\documentclass[aps,prb,twocolumn,eqsecnum,showpacs,groupedaddress]{revtex4}
\usepackage{graphicx,color}
\usepackage{amsmath}
\usepackage{amssymb}
\usepackage{bm}

\renewcommand{\d}{d}
\newcommand{\e}{e}

\renewcommand{\Re}{\mathop{\text{Re}}\nolimits}
\renewcommand{\Im}{\mathop{\text{Im}}\nolimits}
\newcommand{\tr}{\mathop{\text{tr}}\nolimits}
\newcommand{\Tr}{\mathop{\text{Tr}}\nolimits}
\newcommand{\ket}[1]{|{#1}\rangle}
\newcommand{\bra}[1]{\langle{#1}|}

\newcommand{\Det}{\mathop{\text{Det}}\nolimits}

\newcommand{\erfc}{\mathop{\text{erfc}}\nolimits}
\newcommand{\erfi}{\mathop{\text{erfi}}\nolimits}

\definecolor{dgreen}{rgb}{0,0.5,0}

\definecolor{delete}{cmyk}{0.5,0,0,0}

\begin{document}

\title{Quantum Coherence of Electrons Field-Emitted from a Superconductor: Correlations and Entanglement}



\author{Kazuya Yuasa}
\affiliation{Waseda Institute for Advanced Study, Waseda University, Tokyo 169-8050, Japan}



\date[]{September 30, 2009}

\begin{abstract}
The correlations of the electrons field-emitted from a superconductor are fully analyzed, both in space and time.
It is proposed that a coincidence experiment would reveal a positive correlation between the electrons emitted in opposite directions.
The electrons can be entangled and can even violate Bell's inequality. The crucial role played by Andreev's process is scrutinized, analytical formulas are derived for the correlations, and the physics behind the phenomenon is clarified.
\end{abstract}
\pacs{
03.65.Ud,
79.70.+q,
74.45.+c
}


\maketitle

\section{Introduction}
Coherence is a fundamental notion in quantum mechanics.
It is related to the superposition principle and plays a fundamental role whenever intrinsically quantum phenomena take place.
When many identical particles are involved, the global picture can become very rich: quantum coherence, together with quantum statistics, may drive the system into highly nontrivial states, endowed with a variety of interesting features.
The superconducting state of electrons in solids is one of such examples.
It exhibits very interesting physics and has attracted the attention of many researchers from different perspectives.\cite{ref:BCS}

The superconducting state is fully characterized by a set of correlation functions.
If some particles are emitted from the superconductor, they retain some features of the correlations in the source. A superconductor is characterized by its long-range coherence, both in space and time, and by Cooper-pair correlations.
The long coherence can yield a well-monochromatized beam of electrons, which can greatly improve the quality of a microscope.\cite{ref:OshimaNb} 
When a Cooper pair is emitted, its singlet spin state can be a useful resource of entanglement, with applications in quantum information.\cite{ref:Loss,ref:Blatter,ref:SauretFeinbergMartin,ref:Buettiker,ref:Prada,ref:FaoroTaddeiFazio,ref:review-Beenakker,ref:review-Burkard}
It is also interesting to regard the emitted particles as probes of the source.\cite{ref:ProbeCorrelation}
Their correlation functions reflect the features of the source and can behave in remarkable ways.
Positive correlations among electrons have been discussed in the context of superconductors.\cite{ref:PositiveCorr}

A coincidence experiment is a very direct way to detect two-particle correlations.
Recent technological progress has made possible coincidence experiments in a variety of systems.\cite{ref:Bunching-Photon,
ref:Bunching-Atom,
ref:Antibunching-Atom,
ref:Antibunching-Kiesel,
ref:Antibunching-Neutron}
Among these experiments, we shall focus on i) a coincidence experiment in field emission;\cite{ref:Antibunching-Kiesel} ii) the observation\cite{ref:OshimaNb} of the field emission spectrum from a superconductor.\cite{ref:Tunneling-Gadzuk}
A combination of these two experiments can lead to a challenging possibility: a coincidence experiment in the field emission from a superconductor.

The nonlocality of the electrons field-emitted from a superconductor has been recently analyzed.\cite{ref:antibunchingBCS}
This article provides a thorough explanation of this phenomenon, scrutinizing several additional effects and the role played by some genuine superconductivity phenomena, such as the Andreev's process.
The emission process will be dynamically described in the framework of quantum field theory and the beam profile will be naturally prepared by the dynamics itself.\cite{ref:LateralEffects}
It is crucial to work in 3D space to capture the Cooper-pair correlations.
This also enables us to discuss the lateral coherence length.\cite{ref:LateralEffects}
Concise and useful analytical formulas will be derived, which will clarify several facets of the physics behind the phenomenon.
The present analysis also includes correlations in time.
The Andreev process will be shown to play a crucial role for these peculiar correlations and nonlocality.\cite{ref:Loss,ref:Blatter,ref:SauretFeinbergMartin,ref:Buettiker,ref:Prada,ref:FaoroTaddeiFazio,ref:review-Beenakker,ref:review-Burkard}

The article is organized as follows.
The problem is set up and the Hamiltonian of the system is given in Sec.\ \ref{sec:Setup}\@.
The dynamics of the emission is solved and the stationary beam of electrons is obtained in Sec.\ \ref{sec:Emission}\@.
The correlation functions and their spectra are computed in the stationary beam and some analytical formulas, whose structures clarify the underlying physical processes,
are presented in Sec.\ \ref{sec:CorrelationFuncs}\@.
The coincidence spectrum is analyzed in detail in Sec.\ \ref{sec:BunchingAntibunching}, where the effects of the superconductivity of the emitter are disclosed. 
Section \ref{sec:Entanglement} is devoted to the entanglement and nonlocality of the emitted electrons.
The robustness of the correlations due to the superconductivity of the emitter is discussed in Sec.\ \ref{ref:robustness}\@.
Finally, the whole analysis is summarized in Sec.\ \ref{sec:Summary}, and some details are presented in Appendices \ref{app:G}--\ref{app:IntForm}\@.

\section{Setup}
\label{sec:Setup}
\subsection{Hamiltonian}
\begin{figure}[t]
\includegraphics[width=0.30\textwidth]{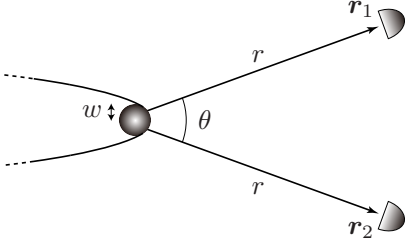}
\caption{Emission of electrons from a superconductor in 3D space and detection of two electrons emitted in different directions.}
\label{fig:Setup}
\end{figure}

Let us start off by discussing electron emission from a superconductor into vacuum. We focus on the emission from a nanotip, see Fig.\ \ref{fig:Setup}. 
To this end, we set up the following Hamiltonian\cite{ref:Tunneling,
ref:Tunneling-Gadzuk,ref:antibunchingBCS,ref:LateralEffects} in 3D space:
\begin{equation}
H=H_0+\lambda H_T,\quad
H_0=H_S+H_V,
\label{eqn:Hamiltonian}
\end{equation}
where
\begin{multline}
H_S
=\int\d^3\bm{r}\,\Biggl[
\sum_{s=\uparrow,\downarrow}\phi_s^\dag(\bm{r})\left(
-\frac{1}{2m}\nabla^2
\right)\phi_s(\bm{r})
\\
{}-W\phi_\downarrow^\dag(\bm{r})
\phi_\uparrow^\dag(\bm{r})
\phi_\uparrow(\bm{r})
\phi_\downarrow(\bm{r})
\Biggr]
\end{multline}
is the Hamiltonian of the superconducting emitter \cite{ref:BCS} and
\begin{equation}
H_V
=\sum_{s=\uparrow,\downarrow}\int\d^3\bm{r}\,
\psi_s^\dag(\bm{r})\left(
-\frac{1}{2m}\nabla^2
\right)\psi_s(\bm{r})
\end{equation}
is that of the electrons propagating in vacuum.
$s$ denotes spin and we set $\hbar=1$.
$\phi_s(\bm{r})$ and $\psi_s(\bm{r})$ are the fermionic field operators satisfying the canonical anticommutation relations
\begin{equation}
\{\phi_s(\bm{r}),\phi_{s'}^\dag(\bm{r}')\}
=\{\psi_s(\bm{r}),\psi_{s'}^\dag(\bm{r}')\}
=\delta_{ss'}\delta^3(\bm{r}-\bm{r}')
\end{equation}
with other anticommutators vanishing. The Coulomb repulsion is neglected.

$H_T$ describes electron tunneling through a potential barrier surrounding the emitting region, and $\lambda$ characterizes the strength of the tunneling transmission.
$H_T$ is given by  \cite{ref:Tunneling,
ref:Tunneling-Gadzuk,ref:antibunchingBCS,ref:LateralEffects}
\begin{equation}
H_T
=\sum_{s=\uparrow,\downarrow}
\int\d^3\bm{p}\int\d^3\bm{k}\,\Bigl(
T_{\bm{p}\bm{k}}c_{\bm{p}s}^\dag a_{\bm{k}s}
+T_{\bm{p}\bm{k}}^*a_{\bm{k}s}^\dag c_{\bm{p}s}
\Bigr),
\label{eqn:HT}
\end{equation}
where $a_{\bm{k}s}$ and $c_{\bm{p}s}$ are the annihilation operators in momentum 
space of the electrons inside and outside the emitter, respectively, and are related to the fields in configuration space by
\begin{subequations}
\label{eqn:Fields}
\begin{align}
\phi_s(\bm{r})
&=\int\frac{\d^3\bm{k}}{\sqrt{(2\pi)^3}}\,
a_{\bm{k}s}\e^{i\bm{k}\cdot\bm{r}},
\displaybreak[0]
\\
\psi_s(\bm{r})
&=\int\frac{\d^3\bm{p}}{\sqrt{(2\pi)^3}}\,
c_{\bm{p}s}\e^{i\bm{p}\cdot\bm{r}}.
\end{align}
\end{subequations}
$T_{\bm{p}\bm{k}}$ are the tunneling matrix elements, for which we take \cite{ref:antibunchingBCS,ref:LateralEffects}
\begin{align}
T_{\bm{p}\bm{k}}
&=\bra{\bm{p}}h(\bm{p})g(\bm{r})\ket{\bm{k}}\nonumber
\displaybreak[0]
\\
&=h(\bm{p})\int\frac{\d^3\bm{r}}{(2\pi)^3}\,
g(\bm{r})\e^{- i(\bm{p}-\bm{k})\cdot\bm{r}}
=h(\bm{p})\tilde{g}(\bm{p}-\bm{k}).
\label{eqn:EmissionMatrix}
\end{align}
That is, an electron with momentum $\bm{k}$ in the emitter is annihilated by  $a_{\bm{k}s}$, filtered by $g(\bm{r})$ and $h(\bm{p})$, and emitted outside with momentum $\bm{p}$ by $c_{\bm{p}s}^\dag$.
The function $g(\bm{r})$ specifies the emitting region and $|h(\bm{p})|^2$ represents the momentum (energy) dependence of the tunneling probability through the potential barrier surrounding the emitting region.
In this article, we consider a simple spherically symmetric setup with
\begin{subequations}
\label{eqn:Setup}
\begin{gather}
g(\bm{r})
=\frac{1}{\sqrt{(2\pi w^2)^3}}
\e^{-r^2/2w^2},\quad
\tilde{g}(\bm{k})
=\frac{1}{(2\pi)^3}
\e^{-k^2w^2/2},\displaybreak[0]\\
h(\bm{p})=\sqrt{\frac{p}{m}}\e^{\varepsilon_p/2E_C},
\end{gather}
\end{subequations}
where $w$ characterizes the size of the emitting region, $\varepsilon_p$ is the energy of an electron, to be given below in (\ref{eqn:Dispersions}), and $E_C$ controls the low-energy cutoff of the tunneling spectrum.
The function $h(\bm{p})$ chosen here might not always accurately describe the tunneling probability through the potential barrier (for which the relevant quantity might be the energy related to the motion normal to the potential surface, instead of the total energy $\varepsilon_p$).
In the following discussion, however, we are interested in the far field, for which the above choice will turn out to be appropriate, since only the momentum normal to the surface is relevant to the far field.

\subsection{Mean-Field Approximation}
Let us introduce
\begin{subequations}
\label{eqn:HamiltonianChemical}
\begin{gather}
\mathcal{H}
=H-\mu N
=\mathcal{H}_0+\lambda H_T,\\
\mathcal{H}_0
=H_0-\mu N,
\end{gather}
\end{subequations}
where
\begin{equation}
N=N_S+N_V
\end{equation}
is the sum of the numbers of electrons inside ($N_S$) and outside ($N_V$) the emitter, defined respectively by
\begin{subequations}
\begin{align}
N_S
&=\sum_{s=\uparrow,\downarrow}\int\d^3\bm{r}\,
\phi_s^\dag(\bm{r})\phi_s(\bm{r}),
\displaybreak[0]
\\
N_V
&=\sum_{s=\uparrow,\downarrow}\int\d^3\bm{r}\,
\psi_s^\dag(\bm{r})\psi_s(\bm{r}).
\end{align}
\end{subequations}
Notice that $N$ is a constant of motion, i.e.\ $[\mathcal{H},N]=0$.
The time-evolution operator can therefore be split as $\e^{- iHt}=\e^{- i\mu Nt}\e^{- i\mathcal{H}t}$, and exponential factors like $\e^{- i\mu t}$ factorize away from the quantities of interest.
Indeed, the Heisenberg operators are factorized as
\begin{equation}
\psi_s(\bm{r},t)
=\e^{ iHt}\psi_s(\bm{r})\e^{- iHt}
=\tilde{\psi}_s(\bm{r},t)
\e^{- i\mu t},
\end{equation}
and so are the correlation functions,
\begin{align}
&\langle\psi_{s_1}^\dag(\bm{r}_1,t_1)\psi_{s_2}(\bm{r}_2,t_2)\rangle
\nonumber\\
&\qquad
=\langle\tilde{\psi}_{s_1}^\dag(\bm{r}_1,t_1)\tilde{\psi}_{s_2}(\bm{r}_2,t_2)\rangle\e^{ i\mu(t_1-t_2)},\ \text{etc.},
\end{align}
where
\begin{equation}
\tilde{\psi}_s(\bm{r},t)
=\e^{ i\mathcal{H}t}\psi_s(\bm{r})\e^{- i\mathcal{H}t}
\end{equation}
describes the dynamics of the field in the picture introduced by the unitary transformation $\e^{ i\mu Nt}$.

We shall work in such a picture, with $\mu$ the Fermi level of the superconducting emitter.
This choice is convenient for the mean-field approximation, which enables one to diagonalize the Hamiltonian $\mathcal{H}_S$ via the Bogoliubov transformation\cite{ref:BCS} 
\begin{equation}
\begin{pmatrix}
\medskip
\displaystyle
a_{\bm{k}\uparrow}\\
\displaystyle
a_{-\bm{k}\downarrow}^\dag
\end{pmatrix}
=\begin{pmatrix}
\medskip
u_k&-v_k\\
v_k^*&u_k^*
\end{pmatrix}
\begin{pmatrix}
\medskip
\displaystyle
\alpha_{\bm{k}\uparrow}\\
\displaystyle
\alpha_{-\bm{k}\downarrow}^\dag
\end{pmatrix}
\label{eqn:BogoliubovTr}
\end{equation}
with
\begin{equation}
\begin{cases}
\medskip
\displaystyle
u_k=\frac{1}{\sqrt{2}}\sqrt{1+\frac{\varepsilon_k}{\omega_k}},\\
\displaystyle
v_k=\frac{\e^{ i\delta}}{\sqrt{2}}\sqrt{1-\frac{\varepsilon_k}{\omega_k}},
\end{cases}
\end{equation}
to get
\begin{subequations}
\label{eqn:FreeHamiltonian}
\begin{align}
\mathcal{H}_S
&=H_S-\mu N_S
=\sum_{s=\uparrow,\downarrow}\int\d^3\bm{k}\,\omega_k
\alpha_{\bm{k}s}^\dag\alpha_{\bm{k}s},
\label{eqn:HS}\displaybreak[0]\\
\mathcal{H}_V
&=H_V-\mu N_V
=\sum_{s=\uparrow,\downarrow}
\int\d^3\bm{p}\,\varepsilon_pc_{\bm{p}s}^\dag c_{\bm{p}s},
\label{eqn:HV}
\end{align}
\end{subequations}
where 
\begin{equation}
\varepsilon_p=\frac{p^2}{2m}-\mu,
\qquad
\omega_k=\sqrt{\varepsilon_k^2+|\Delta|^2}
\label{eqn:Dispersions}
\end{equation}
are the energies of an emitted electron in vacuum and of a quasiparticle excitation in the superconducting emitter, respectively, measured relative to the Fermi level of the emitter, and
\begin{equation}
\Delta=W\langle\phi_\uparrow(\bm{r})\phi_\downarrow(\bm{r})\rangle=|\Delta|\e^{ i\delta}
\end{equation}
is the gap parameter of the superconductor.
Throughout this article, $k_F=\sqrt{2m\mu}$ and $\lambda_F=2\pi/k_F$ are the Fermi momentum and the Fermi wavelength, respectively.

The state of the superconductor is characterized by the Fermi distribution of the quasiparticle excitations,
\begin{equation}
\langle\alpha_{\bm{k}s}^\dag\alpha_{\bm{k}'s'}\rangle
=f(\omega_k)\delta_{ss'}\delta^3(\bm{k}-\bm{k}')
\label{eqn:QuasiPartExcitations}
\end{equation}
with
\begin{equation}
f(\omega_k)=\frac{1}{\e^{\omega_k/k_BT}+1},
\end{equation}
where $T$ is the temperature of the emitter and $k_B$ the Boltzmann constant.
The spatial extension of a Cooper pair is characterized by the correlation length of the two-point correlation function $\langle\phi_\uparrow(\bm{r}_1)\phi_\downarrow(\bm{r}_2)\rangle$ in the superconductor, which is given (at zero temperature) by Pippard's length\cite{ref:BCS}
\begin{equation}
\xi=\frac{k_F}{\pi m|\Delta|}=\frac{2\mu}{\pi k_F|\Delta|}.
\label{eqn:Pippard}
\end{equation}

\subsection{Coincidence}
At the initial time $t=0$, the emitter is in the superconducting state at a given temperature $T$ and the outside of the emitter is vacuum.
Such an initial state is a product state of the superconducting state characterized by (\ref{eqn:QuasiPartExcitations}) and the vacuum state for $c_{\bm{p}s}$.
Starting from this initial condition, the electrons start to tunnel out of the emitter into vacuum.
The emission process is dynamically described according to the Hamiltonian (\ref{eqn:HamiltonianChemical}) with (\ref{eqn:HT}) and (\ref{eqn:FreeHamiltonian}) in the mean-field approximation.
After a certain transient period, the emission approaches a nonequilibrium steady state (NESS).\cite{ref:NESS}

We shall count coincidences, in the NESS prepared in this way, between two detectors located at
\begin{equation}
\bm{r}_{1,2}=\bm{(}\pm r\sin(\theta/2),0,r\cos(\theta/2)\bm{)},
\end{equation}
at the same distance from the emitter but in different directions, as depicted in Fig.\ \ref{fig:Setup}.
The numbers of counts of one- and two-particle detections, irrespectively of the spin state, are proportional to
\begin{equation}
\rho^{(1)}(\bm{r},t)
=\sum_{s=\uparrow,\downarrow}
\langle\psi_s^\dag(\bm{r},t)\psi_s(\bm{r},t)\rangle,
\label{eqn:OneDistriDef}
\end{equation}
\begin{align}
&\rho^{(2)}(\bm{r}_2,t_2;\bm{r}_1,t_1)
\nonumber\\
&\ %
=\sum_{s_1,s_2=\uparrow,\downarrow}
\langle\psi_{s_1}^\dag(\bm{r}_1,t_1)\psi_{s_2}^\dag(\bm{r}_2,t_2)
\psi_{s_2}(\bm{r}_2,t_2)\psi_{s_1}(\bm{r}_1,t_1)\rangle
\nonumber
\\[-2mm]
&\hspace*{71mm}
(t_2\ge t_1),
\label{eqn:TwoDistriDef}
\end{align}
respectively, where the average is taken over the initial state.
The deviation of the normalized two-particle distribution function
\begin{equation}
Q(\bm{r}_2,t_2;\bm{r}_1,t_1)
=\frac{
\rho^{(2)}(\bm{r}_2,t_2;\bm{r}_1,t_1)
}{
\rho^{(1)}(\bm{r}_2,t_2)
\rho^{(1)}(\bm{r}_1,t_1)
}\quad
(t_2\ge t_1)
\label{eqn:NormalizedCorrelation}
\end{equation}
from unity reveals correlations between the two electrons.

\section{Dynamics of Emission}
\label{sec:Emission}
Let us solve the dynamics of the emission. 
The Heisenberg equations of motion for $\tilde{c}_{\bm{p}s}(t)=\e^{ i\mathcal{H}t}c_{\bm{p}s}\e^{- i\mathcal{H}t}$, etc., read
\begin{subequations}
\label{eqn:HeisenbergEq}
\begin{gather}
\frac{\d}{\d t}
\begin{pmatrix}
\medskip
\tilde{c}_{\bm{p}\uparrow}\\
\tilde{c}_{\bm{p}\downarrow}^\dag
\end{pmatrix}
=- i\mathcal{E}_p
\begin{pmatrix}
\medskip
\tilde{c}_{\bm{p}\uparrow}\\
\tilde{c}_{\bm{p}\downarrow}^\dag
\end{pmatrix}
- i\lambda\int\d^3\bm{k}\,\mathcal{T}_{\bm{p}\bm{k}}
\begin{pmatrix}
\medskip
\tilde{\alpha}_{\bm{k}\uparrow}\\
\tilde{\alpha}_{\bm{k}\downarrow}^\dag
\end{pmatrix},
\displaybreak[0]
\\
\frac{\d}{\d t}
\begin{pmatrix}
\medskip
\tilde{\alpha}_{\bm{k}\uparrow}\\
\tilde{\alpha}_{\bm{k}\downarrow}^\dag
\end{pmatrix}
=- i\varOmega_k
\begin{pmatrix}
\medskip
\tilde{\alpha}_{\bm{k}\uparrow}\\
\tilde{\alpha}_{\bm{k}\downarrow}^\dag
\end{pmatrix}
- i\lambda\int\d^3\bm{p}\,\mathcal{T}_{\bm{p}\bm{k}}^\dag
\begin{pmatrix}
\medskip
\tilde{c}_{\bm{p}\uparrow}\\
\tilde{c}_{\bm{p}\downarrow}^\dag
\end{pmatrix},
\end{gather}
\end{subequations}
where
\begin{gather}
\mathcal{E}_p
=\begin{pmatrix}
\medskip
\varepsilon_p&0\\
0&-\varepsilon_p
\end{pmatrix},\qquad
\varOmega_k
=\begin{pmatrix}
\medskip
\omega_k&0\\
0&-\omega_k
\end{pmatrix},
\displaybreak[0]
\\
\mathcal{T}_{\bm{p}\bm{k}}
=\begin{pmatrix}
\medskip
T_{\bm{p}\bm{k}}u_k&-T_{\bm{p}(-\bm{k})}v_k\\
-T_{\bm{p}(-\bm{k})}^*v_k^*&-T_{\bm{p}\bm{k}}^*u_k^*
\end{pmatrix}.
\end{gather}
This set of equations is solved by means of the Laplace transform to yield
\begin{align}
&\begin{pmatrix}
\medskip
\tilde{c}_{\bm{p}\uparrow}(t)\\
\tilde{c}_{\bm{p}\downarrow}^\dag(t)
\end{pmatrix}
=\int\d^3\bm{p}'\,\mathcal{G}_{\bm{p}\bm{p}'}(t)
\begin{pmatrix}
\medskip
c_{\bm{p}'\uparrow}\\
c_{\bm{p}'\downarrow}^\dag
\end{pmatrix}
\nonumber
\displaybreak[0]
\\
&\qquad\quad
{}- i\lambda
\int_0^t\d t'\int\d^3\bm{k}\,
[\mathcal{G}(t-t')\mathcal{T}]_{\bm{p}\bm{k}}
\e^{- i\varOmega_kt'}
\begin{pmatrix}
\medskip
\alpha_{\bm{k}\uparrow}\\
\alpha_{\bm{k}\downarrow}^\dag
\end{pmatrix},
\label{eqn:HeisenbergSol}
\end{align}
where 
$\mathcal{G}_{\bm{p}\bm{p}'}(t)$ is given by the inverse Laplace transform 
\begin{equation}
\mathcal{G}_{\bm{p}\bm{p}'}(t)
=\int_{C_B}\frac{\d s}{2\pi i }\,
\hat{\mathcal{G}}_{\bm{p}\bm{p}'}(s)\e^{st}
\label{eqn:Gdef}
\end{equation}
of
\begin{gather}
\hat{\mathcal{G}}_{\bm{p}\bm{p}'}^{-1}(s)
=(s+ i\mathcal{E}_p)\delta^3(\bm{p}-\bm{p}')
+\lambda^2\hat{\mathcal{K}}_{\bm{p}\bm{p}'}(s),
\label{eqn:Ginv}
\displaybreak[0]\\
\hat{\mathcal{K}}_{\bm{p}\bm{p}'}(s)
=\int\d^3\bm{k}\,\mathcal{T}_{\bm{p}\bm{k}}
\frac{1}{s+ i\varOmega_k}
\mathcal{T}_{\bm{p}'\bm{k}}^\dag,
\label{eqn:K}
\end{gather}
with $C_B$ the Bromwich path running parallel at the right of the imaginary axis of $s$, and $[\mathcal{G}(t-t')\mathcal{T}]_{\bm{p}\bm{k}}=\int\d^3\bm{p}'\,\mathcal{G}_{\bm{p}\bm{p}'}(t-t')\mathcal{T}_{\bm{p}'\bm{k}}$ like a matrix product.

In order to obtain the NESS, it is convenient to move to the interaction picture defined by $\bar{c}_{\bm{p}s}(t)=\e^{- i\mathcal{H}_0t_0}\tilde{c}_{\bm{p}s}(t)\e^{ i\mathcal{H}_0t_0}$, since the initial state is invariant under this transformation.
In this picture, $\bar{\mathcal{G}}_{\bm{p}\bm{p}'}(t)=\mathcal{G}_{\bm{p}\bm{p}'}(t)\e^{ i\mathcal{E}_{p'}t_0}$ asymptotically behaves, for $t,t_0\to\infty$, keeping $t-t_0$ finite, as (Appendix \ref{app:G})
\begin{widetext}
\begin{equation}
\bar{\mathcal{G}}_{\bm{p}\bm{p}'}(t)
\to
\e^{- i\mathcal{E}_p(t-t_0)}\delta^3(\bm{p}-\bm{p}')
-\lambda^2
\begin{pmatrix}
\medskip
\frac{
\hat{K}_{\bm{p}\bm{p}'}^{11}(- i\varepsilon_{p'}+0^+)
}{ i(\varepsilon_p-\varepsilon_{p'})+0^+}
\e^{- i\varepsilon_{p'}(t-t_0)}
&\frac{
\hat{K}_{\bm{p}\bm{p}'}^{12}( i\varepsilon_{p'}+0^+)
}{ i(\varepsilon_p+\varepsilon_{p'})+0^+}
\e^{ i\varepsilon_{p'}(t-t_0)}
\\
\frac{
\hat{K}_{\bm{p}\bm{p}'}^{21}(- i\varepsilon_{p'}+0^+)
}{- i(\varepsilon_p+\varepsilon_{p'})+0^+}
\e^{- i\varepsilon_{p'}(t-t_0)}
&\frac{
\hat{K}_{\bm{p}\bm{p}'}^{22}( i\varepsilon_{p'}+0^+)
}{- i(\varepsilon_p-\varepsilon_{p'})+0^+}
\e^{ i\varepsilon_{p'}(t-t_0)}
\end{pmatrix}
+O(\lambda^4),
\label{eqn:Gasymp}
\end{equation}
where
\begin{equation}
\hat{\mathcal{K}}_{\bm{p}\bm{p}'}(s)
=\begin{pmatrix}
\medskip
\hat{K}_{\bm{p}\bm{p}'}^{11}(s)&
\hat{K}_{\bm{p}\bm{p}'}^{12}(s)\\
\hat{K}_{\bm{p}\bm{p}'}^{21}(s)&
\hat{K}_{\bm{p}\bm{p}'}^{22}(s)
\end{pmatrix}
=\int\d^3\bm{k}
\begin{pmatrix}
\medskip
T_{\bm{p}\bm{k}}T_{\bm{p}'\bm{k}}^*\left(
\frac{|u_k|^2}{s+ i\omega_k}
+\frac{|v_k|^2}{s- i\omega_k}
\right)&
T_{\bm{p}\bm{k}}T_{\bm{p}'(-\bm{k})}u_kv_k\left(
\frac{1}{s- i\omega_k}
-\frac{1}{s+ i\omega_k}
\right)\\
T_{\bm{p}\bm{k}}^*T_{\bm{p}'(-\bm{k})}^*u_k^*v_k^*\left(
\frac{1}{s- i\omega_k}
-\frac{1}{s+ i\omega_k}
\right)&
T_{\bm{p}\bm{k}}^*T_{\bm{p}'\bm{k}}\left(
\frac{|u_k|^2}{s- i\omega_k}
+\frac{|v_k|^2}{s+ i\omega_k}
\right)
\end{pmatrix}.
\end{equation}
Note that we need to retain up to $O(\lambda^2)$ to collect the lowest-order contributions to the
correlation function (\ref{eqn:TwoDistriDef}), which is
$O(\lambda^4)$.
Up to this order, the Heisenberg operators
(\ref{eqn:HeisenbergSol}) in the NESS in the interaction
picture read
\begin{equation}
\begin{pmatrix}
\medskip
\bar{c}_{\bm{p}\uparrow}(t)\\
\bar{c}_{\bm{p}\downarrow}^\dag(t)
\end{pmatrix}
\to\int\d^3\bm{p}'\,\bar{\mathcal{G}}_{\bm{p}\bm{p}'}(t)
\begin{pmatrix}
\medskip
c_{\bm{p}'\uparrow}\\
c_{\bm{p}'\downarrow}^\dag
\end{pmatrix}
-\lambda\int\d^3\bm{k}
\begin{pmatrix}
\medskip
\frac{T_{\bm{p}\bm{k}}u_k}{\varepsilon_p-\omega_k-i0^+}&
-\frac{T_{\bm{p}(-\bm{k})}v_k}{\varepsilon_p+\omega_k-i0^+}\\
\frac{T_{\bm{p}(-\bm{k})}^*v_k^*}{\varepsilon_p+\omega_k+i0^+}&
\frac{T_{\bm{p}\bm{k}}^*u_k^*}{\varepsilon_p-\omega_k+i0^+}
\end{pmatrix}
\e^{- i\varOmega_k(t-t_0)}
\begin{pmatrix}
\medskip
\alpha_{\bm{k}\uparrow}\\
\alpha_{\bm{k}\downarrow}^\dag
\end{pmatrix}
+O(\lambda^3).
\end{equation}

\section{Correlation Functions}
\label{sec:CorrelationFuncs}
\subsection{Structures of the Correlation Functions}
\label{sec:Structure}
The correlation functions of the emitted electrons in the NESS are now readily computed.
Let us start with two-point correlation functions.
In the limit $t_1,t_2\to\infty$, keeping $t_1-t_2$ finite, one gets, up to $O(\lambda^2)$,
\begin{align}
\gamma(\bm{r}_1,t_1;\bm{r}_2,t_2)
&
=\langle
\tilde{\psi}_\uparrow^\dag(\bm{r}_1,t_1)
\tilde{\psi}_\uparrow(\bm{r}_2,t_2)
\rangle
=\langle
\tilde{\psi}_\downarrow^\dag(\bm{r}_1,t_1)
\tilde{\psi}_\downarrow(\bm{r}_2,t_2)
\rangle
\nonumber
\displaybreak[0]
\\
&
\to\lambda^2\sum_{\sigma=\pm}
\int\d^3\bm{k}\,f(\sigma\omega_k)
\varphi_{\bm{k}\sigma}^*(\bm{r}_1,t_1)
\varphi_{\bm{k}\sigma}(\bm{r}_2,t_2),
\label{eqn:Gamma}
\end{align}
\begin{align}
\chi(\bm{r}_1,t_1;\bm{r}_2,t_2)
&
=\langle
\tilde{\psi}_\uparrow(\bm{r}_1,t_1)
\tilde{\psi}_\downarrow(\bm{r}_2,t_2)
\rangle
=-\langle
\tilde{\psi}_\downarrow(\bm{r}_1,t_1)
\tilde{\psi}_\uparrow(\bm{r}_2,t_2)
\rangle
\nonumber
\displaybreak[0]
\\
&
\to\chi_0(\bm{r}_1,t_1;\bm{r}_2,t_2)
+\chi_\text{th}(\bm{r}_1,t_1;\bm{r}_2,t_2),
\label{eqn:Chi}
\end{align}
where
\begin{equation}
\chi_\text{th}(\bm{r}_1,t_1;\bm{r}_2,t_2)
=-\lambda^2\sum_{\sigma=\pm}
\int\d^3\bm{k}\,f(\omega_k)
\varphi_{(-\bm{k})(-\sigma)}(\bm{r}_1,t_1)
\varphi_{\bm{k}\sigma}(\bm{r}_2,t_2),
\label{eqn:ChiTh}
\end{equation}
\begin{align}
&\chi_0(\bm{r}_1,t_1;\bm{r}_2,t_2)
\nonumber
\\
&\ %
=\lambda^2\int\d^3\bm{k}\,u_kv_k
\int\frac{\d^3\bm{p}_1}{\sqrt{(2\pi)^3}}
\frac{\d^3\bm{p}_2}{\sqrt{(2\pi)^3}}
\frac{T_{\bm{p}_1\bm{k}}T_{\bm{p}_2(-\bm{k})}
}{\varepsilon_{p_1}+\varepsilon_{p_2}-i0^+}
\left(
\frac{1}{\varepsilon_{p_1}+\omega_k-i0^+}
+\frac{1}{\varepsilon_{p_2}+\omega_k-i0^+}
\right)
\e^{i\bm{p}_1\cdot\bm{r}_1}\e^{i\bm{p}_2\cdot\bm{r}_2}
\e^{- i\varepsilon_{p_1}(t_1-t_2)}
\nonumber
\\
&\ \quad
{}-\lambda^2\int\d^3\bm{k}\,
u_kv_k
\int\frac{\d^3\bm{p}_1}{\sqrt{(2\pi)^3}}
\frac{\d^3\bm{p}_2}{\sqrt{(2\pi)^3}}
\frac{T_{\bm{p}_1\bm{k}}T_{\bm{p}_2(-\bm{k})}\e^{i\bm{p}_1\cdot\bm{r}_1}\e^{i\bm{p}_2\cdot\bm{r}_2}}{(\varepsilon_{p_1}-\omega_k-i0^+)(\varepsilon_{p_2}+\omega_k-i0^+)}
(\e^{- i\varepsilon_{p_1}(t_1-t_2)}-\e^{- i\omega_k(t_1-t_2)}),
\label{eqn:Andreev}
\end{align}
\end{widetext}
while all other two-point correlation functions vanish.
In these expressions,
\begin{subequations}
\label{eqn:WaveFunction}
\begin{align}
\varphi_{\bm{k}+}(\bm{r},t)
&=\int\frac{\d^3\bm{p}}{\sqrt{(2\pi)^3}\, i }
\frac{T_{\bm{p}\bm{k}}u_k}{\varepsilon_p-\omega_k-i0^+}
\e^{ i(\bm{p}\cdot\bm{r}-\omega_kt)},
\displaybreak[0]\\
\varphi_{\bm{k}-}(\bm{r},t)
&=\int\frac{\d^3\bm{p}}{\sqrt{(2\pi)^3}\, i }
\frac{T_{\bm{p}\bm{k}}v_k}{\varepsilon_p+\omega_k-i0^+}
\e^{ i(\bm{p}\cdot\bm{r}+\omega_kt)}
\end{align}
\end{subequations}
are the wave functions of the emitted electrons in the vacuum originating from the quasiparticle excitations above and below the Fermi level, respectively, and they propagate asymptotically as spherical waves (Appendix \ref{app:AsympWave})
\begin{subequations}
\label{eqn:AsympWaveFunction}
\begin{align}
\varphi_{\bm{k}+}(\bm{r},t)
&\simeq m\sqrt{2\pi}\,T_{(p_+\hat{\bm{r}})\bm{k}}u_k
\frac{1}{ i r}\e^{ i(p_+r-\omega_kt)},
\displaybreak[0]\\
\varphi_{\bm{k}-}(\bm{r},t)
&\simeq m\sqrt{2\pi}\,\theta(\mu-\omega_k)
T_{(p_-\hat{\bm{r}})\bm{k}}v_k
\frac{1}{ i r}\e^{ i(p_-r+\omega_kt)}
\end{align}
\end{subequations}
for $k_Fr\gg1$, where
\begin{equation}
p_\pm=p(\pm\omega_k),\quad
p(E)=\sqrt{2m(\mu+E)}
\label{eqn:Ppm}
\end{equation}
are the momenta of the emitted electrons corresponding to the energies $\pm\omega_k$, and $\theta(x)$ is the Heaviside theta function.

The correlation function $\gamma$ in (\ref{eqn:Gamma}) is a mixture of the wave functions $\varphi_{\bm{k}\pm}(\bm{r},t)$ over the Fermi distribution $f(\pm\omega_k)$ in the emitter, with the relevant densities of states of the quasiparticle excitations, $|u_k|^2$ and $|v_k|^2$.
The quasiparticles are emitted outside and the electrons propagate with the wave functions $\varphi_{\bm{k}\pm}(\bm{r},t)$.
The correlation function $\gamma$ is essentially a one-particle density matrix and describes the one-particle state of the emitted electrons.

The other correlation function $\chi$ in (\ref{eqn:Chi})--(\ref{eqn:Andreev}), on the other hand, describes pair emission.
It contains the product $u_kv_k\,(=\Delta/2\omega_k)$ and is responsible for the effects of the Cooper-pair correlation in the superconducting emitter.
Indeed, $\chi_0$ describes the emission of a couple at zero temperature, from the Cooper pairs, and takes into account the Andreev process.\cite{
ref:HekkingNazarov,
ref:Loss,
ref:Prada} 
Notice that the last contribution in (\ref{eqn:Andreev}) vanishes on the energy shell $\varepsilon_{p_1}=\omega_k$.
The main contribution is the first addendum and it is exactly the amplitude for the emission of a couple [see (\ref{eqn:AndreevAmp}) in Appendix \ref{app:Andreev}].
The contribution of the thermal excitations at finite temperature is taken into account by $\chi_\text{th}$, which vanishes at zero temperature.

Since the solution of the Heisenberg equations of motion, (\ref{eqn:HeisenbergSol}), is linear in the fermionic operators, the one- and two-particle distribution functions, (\ref{eqn:OneDistriDef}) and (\ref{eqn:TwoDistriDef}) are both given in terms of the two-point correlation functions: the former is given by
\begin{equation}
\rho^{(1)}(\bm{r},t)
=2\gamma(\bm{r},t;\bm{r},t),
\label{eqn:OneDistri}
\end{equation}
while the latter is cast through Wick's theorem into
\begin{align}
&\rho^{(2)}(\bm{r}_2,t_2;\bm{r}_1,t_1)
\nonumber
\\
&\quad
=4\gamma(\bm{r}_2,t_2;\bm{r}_2,t_2)
\gamma(\bm{r}_1,t_1;\bm{r}_1,t_1)
\nonumber
\\
&\quad\quad
-2|\gamma(\bm{r}_2,t_2;\bm{r}_1,t_1)|^2
+2|\chi(\bm{r}_2,t_2;\bm{r}_1,t_1)|^2.
\label{eqn:TwoDistri}
\end{align}
The normalized two-particle distribution (\ref{eqn:NormalizedCorrelation}) is therefore given by
\begin{equation}
Q(2;1)
=1-\frac{|\gamma(2;1)|^2}{2\gamma(2;2)\gamma(1;1)}
+\frac{|\chi(2;1)|^2}{2\gamma(2;2)\gamma(1;1)},
\label{eqn:NormalizedCorrelationST}
\end{equation}
where the arguments $(\bm{r}_2,t_2;\bm{r}_1,t_1)$ are abbreviated to $(2;1)$.

The second term of the normalized coincidence (\ref{eqn:NormalizedCorrelationST}) is responsible for the antibunching of electrons (reduction of the coincidence rate when the two detectors are close together and the detection delay time is small), while the third gives a positive contribution to the coincidences.
It is instructive to rewrite the two-particle distribution (\ref{eqn:TwoDistri}) in terms of the two-particle wave functions: by substituting the two-point correlation function $\gamma$ given in (\ref{eqn:Gamma}), i.e.\ the contributions of the quasiparticle excitations, the first two terms in (\ref{eqn:TwoDistri}) read\cite{ref:LateralEffects}
\begin{widetext}
\begin{equation}
\rho^{(2)}(2;1)
=\sum_{\sigma_2,\sigma_1=\pm}
\int\d^3\bm{k}_2
\int\d^3\bm{k}_1\,
f(\sigma_2\omega_{k_2})f(\sigma_1\omega_{k_1})
\,\Bigl(
3\left|
\Psi_{\bm{k}_2\sigma_2,\bm{k}_1\sigma_1}^{(-)}(2;1)
\right|^2
+\left|
\Psi_{\bm{k}_2\sigma_2,\bm{k}_1\sigma_1}^{(+)}(2;1)
\right|^2
\Bigr)
+2|\chi(2;1)|^2,
\label{eqn:TripSingPair}
\end{equation}
\end{widetext}
where 
\begin{align}
&\Psi_{\bm{k}_1\sigma_1,\bm{k}_2\sigma_2}^{(\pm)}(1;2)
\nonumber\\
&\qquad
=\frac{1}{\sqrt{2}}\,\Bigl(
\varphi_{\bm{k}_1\sigma_1}(1)
\varphi_{\bm{k}_2\sigma_2}(2)
\pm\varphi_{\bm{k}_2\sigma_2}(1)
\varphi_{\bm{k}_1\sigma_1}(2)
\Bigr)
\end{align}
are the symmetrized/antisymmetrized two-particle wave functions of the emitted electrons.
This expression clarifies that 3/4 of the contributions of the quasiparticle excitations are provided by the antisymmetric wave function $\Psi^{(-)}$, while the remaining 1/4 is given by the symmetric one $\Psi^{(+)}$.
Recall that the state of the fermions as a whole should be antisymmetric.
The symmetric wave function in space therefore corresponds to the antisymmetric state in spin, and vice versa.
As already mentioned, $\gamma$ describes the emission of the single particles.
The electrons in the emitter are not spin polarized when one looks at the single electrons and they appear to be equally emitted from the triplet (antisymmetric in space) and singlet (symmetric in space) spin states.
This just gives the background: the interplay between these contributions yields antibunching \cite{ref:Antibunching-Neutron,ref:LateralEffects} but does not reveal the effect of the Cooper-pair correlation in the emitter.

The Cooper pairs do not significantly contribute to antibunching, but yield the positive correlation, i.e.\ the last contribution in (\ref{eqn:NormalizedCorrelationST}).
Let us look at the Andreev amplitude, the first contribution to $\chi_0$ in (\ref{eqn:Andreev}).
This expression entails that one electron in a pair originates with momentum $\bm{k}$ in the emitter, while the other with momentum $-\bm{k}$.
Notice here that the integrand is symmetric under exchange between $\bm{k}$ and $-\bm{k}$ and the two alternatives are symmetrically superposed.
This symmetric superposition reflects the fact that the Cooper pairs in the emitter form singlet couples, antisymmetric in spin.
This symmetry in the amplitude with respect to the spatial degrees of freedom gives rise to \textit{bunching}: two electrons bunch when they originate from momenta with the same magnitudes but oriented in opposite directions.
This is the origin of the positive contribution in (\ref{eqn:NormalizedCorrelationST}).

Positive correlation of electrons due to superconductivity has been discussed in the literature, in the current fluctuations in transport setups.\cite{ref:PositiveCorr}
For instance, it is shown that the cross-correlation of the currents at two different normal leads connected to a superconductor can be both negative and positive, while it is quite generally negative in normal devices.
It is pointed out that the positive correlation is due to the processes involving ``crossed Andreev reflections,'' which provide transport of particles with opposite charges from one normal lead to the other via the superconductor (an electron in one normal lead is converted into a hole in the other normal lead, and vice versa),\cite{ref:Loss,ref:Blatter,ref:SauretFeinbergMartin,ref:Buettiker,ref:Prada,ref:FaoroTaddeiFazio,ref:review-Burkard,ref:PositiveCorr,ref:CrossedAndreevTh,ref:CrossedAndreevExp} and the sign of the overall cross-correlation is determined by the  competition between the negative contributions due to the normal scattering and the positive contributions due to the Andreev reflections.
In the present field-emission setup, there is no ``hole'' in vacuum, but the crossed Andreev reflection is equivalent to the emission of electrons from a common Cooper pair toward different detectors.
In the present analysis, it is clear from the formula, as just explained above, that the positive correlation is due to bunching as a result of the symmetric wave function of a couple of electrons emitted from a Cooper pair, reflecting its singlet spin state.
In addition, it is shown below that the negative and positive contributions appear separately in the present field-emission setup, and the pure positive correlation is observed: one need not argue the competition between the negative and positive contributions.

The amplitude for the Andreev emission (\ref{eqn:Andreev}) is also symmetric under the exchange between $\bm{r}_1$ and $\bm{r}_2$.
This symmetry reveals that the emitted electron pairs are also in the singlet state.
This point is explicit in the expression (\ref{eqn:AndreevAmp}).
The entanglement of the emitted pair is one of the main subjects of this article and is explored below.

It is interesting to note that such emission of the singlet pair is not simply the direct emission of the Cooper pair since virtual processes are involved, as is clear from the expression in (\ref{eqn:AndreevAmp}).

\subsection{Spectral Representations}
Let us look at the energy spectra of the correlation functions, $\Gamma(\bm{r}_1,\bm{r}_2;E)$, $\mathcal{X}_\text{th}(\bm{r}_1,\bm{r}_2;E)$, and $\mathcal{X}_0(\bm{r}_1,\bm{r}_2;E)$, which are related to the correlation functions $\gamma(\bm{r}_1,t_1;\bm{r}_2,t_2)$, $\chi_\text{th}(\bm{r}_1,t_1;\bm{r}_2,t_2)$, and $\chi_0(\bm{r}_1,t_1;\bm{r}_2,t_2)$ in (\ref{eqn:Gamma})--(\ref{eqn:Andreev}), respectively, through the Fourier transformations,
\begin{widetext}
\begin{equation}
\gamma(\bm{r}_1,t_1;\bm{r}_2,t_2)
=\int_{-\infty}^\infty\frac{\d E}{2\pi}\,
\Gamma(\bm{r}_1,\bm{r}_2;E)\e^{ i E(t_1-t_2)},
\end{equation}
etc. 
By plugging the asymptotic forms of the wave functions in (\ref{eqn:AsympWaveFunction}) together with the tunneling matrix (\ref{eqn:EmissionMatrix})--(\ref{eqn:Setup}), the energy spectra in the far field region ($k_Fr\gg1$) in the NESS ($t\to\infty$) read (Appendix \ref{app:AsympWave})
\begin{subequations}
\label{eqn:Spectra}
\begin{gather}
\Gamma(\bm{r}_1,\bm{r}_2;E)
\simeq2\pi A
\theta(|E|-|\Delta|)
\frac{|E|}{\sqrt{E^2-|\Delta|^2}}Z(E)f(E)\e^{E/E_C},
\label{eqn:GammaE}
\displaybreak[0]
\\
\mathcal{X}_\text{th}(\bm{r}_1,\bm{r}_2;E)
\simeq2\pi A
\theta(|E|-|\Delta|)\frac{\Delta}{\sqrt{E^2-|\Delta|^2}}
Z_\text{th}(E)f(|E|)\e^{ i[p(E)+p(-E)]r},
\label{eqn:ChiThE}
\displaybreak[0]
\\
\mathcal{X}_0(\bm{r}_1,\bm{r}_2;E)
\simeq\pi A
\frac{\Delta}{\sqrt{E^2-|\Delta|^2}}Z_0(E)\e^{ i[p(E)+p(-E)]r}
\label{eqn:Chi0E}
\end{gather}
\end{subequations}
for $|E|<\mu$, where $A=\lambda^2m^2/2(2\pi)^4w^2r^2$, and
\begin{subequations}
\label{eqn:Zs}
\begin{align}
&Z(E)
=\frac{1}{2\cos(\theta/2)}\sum_{\sigma=\pm}
\left(
1+\sigma\frac{\sqrt{E^2-|\Delta|^2}}{E}
\right)
\e^{-w^2[p(E)-k_\sigma(E)]^2}
\left(
\e^{-4w^2p(E)k_\sigma(E)\sin^2(\theta/4)}
-\e^{-4w^2p(E)k_\sigma(E)\cos^2(\theta/4)}
\right),
\label{eqn:Z}
\displaybreak[0]
\\
&Z_\text{th}(E)
=\frac{\sqrt{p(E)p(-E)}}{2q(E)}
\e^{-w^2[k_F^2-q^2(E)]}
\sum_{\sigma=\pm}
\left(
\e^{-w^2[k_\sigma(E)-q(E)]^2}
-\e^{-w^2[k_\sigma(E)+q(E)]^2}
\right),
\label{eqn:Zth}
\displaybreak[0]
\\
&Z_0(E)=Z_\text{th}(E)+\delta Z_0(E),
\nonumber
\displaybreak[0]
\\
&\qquad\qquad
\delta Z_0(E)
=i\frac{\sqrt{p(E)p(-E)}}{2q(E)}
\e^{-w^2[k_F^2-q^2(E)]}
\sum_{\sigma=\pm}
\sigma\,\Bigl(
\Xi\bm{(}
w[k_\sigma(E)-q(E)]
\bm{)}
-\Xi\bm{(}
w[k_\sigma(E)+q(E)]
\bm{)}
\Bigr)
\label{eqn:Z0}
\end{align}
\end{subequations}
\end{widetext}
with $p(E)$ defined in (\ref{eqn:Ppm}) and
\begin{align}
&k_\pm(E)
=\sqrt{2m\,\Bigl(\mu\pm\sqrt{E^2-|\Delta|^2}\Bigr)},
\\
&q(E)
=\sqrt{[k_F^2-p(E)p(-E)\cos\theta]/2}.
\end{align}
Note that
\begin{equation}
\Xi(z)=\e^{-z^2}\erfi(z)
\end{equation}
is a slowly varying function of $z$ and that we have chosen the branch of the square root $\sqrt{E^2-|\Delta|^2}= i\sqrt{|\Delta|^2-E^2}$ for $|E|<|\Delta|$.

\begin{figure}[b]
\includegraphics[width=0.45\textwidth]{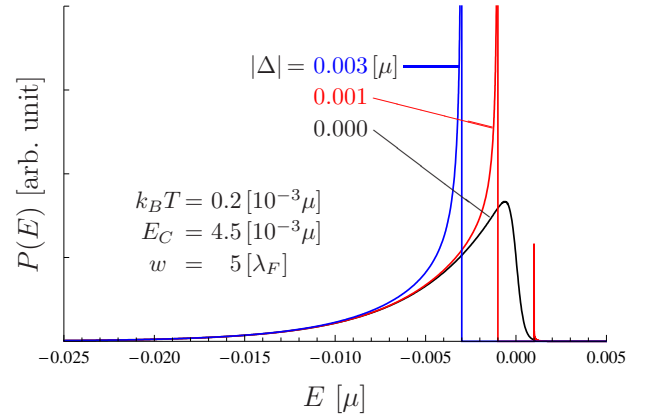}
\caption{(Color online) Energy spectrum of the emitted electrons, (\ref{eqn:Spectrum}) with (\ref{eqn:GammaE}) and (\ref{eqn:Z}).
}
\label{fig:EnergySpectrum}
\end{figure}

The energy spectrum of the emitted electrons,
\begin{equation}
P(E)\propto4\pi r^2\Gamma(\bm{r},\bm{r};E),
\label{eqn:Spectrum}
\end{equation}
is shown in Fig.\ \ref{fig:EnergySpectrum}, where on the basis of the formulas (\ref{eqn:GammaE}) and (\ref{eqn:Z}).
The low-energy cutoff reflects the energy dependence of the tunneling probability: electrons with a lower energy feel a higher and thicker potential barrier and tunneling is suppressed.
In the present model with (\ref{eqn:Setup}), the low-energy slope is characterized by the factor $\e^{E/E_C}$.
The high-energy cutoff, on the other hand, is due to the Fermi distribution $f(E)$ and is therefore mainly controlled by the temperature.
The band width of the spectrum is controlled by the factor $f(E)\e^{E/E_C}$ and the spectrum spans the range
\begin{equation}
-E_C\lesssim E\lesssim(1/k_BT-1/E_C)^{-1}.
\end{equation}

\begin{figure}[t]
\includegraphics[width=0.45\textwidth]{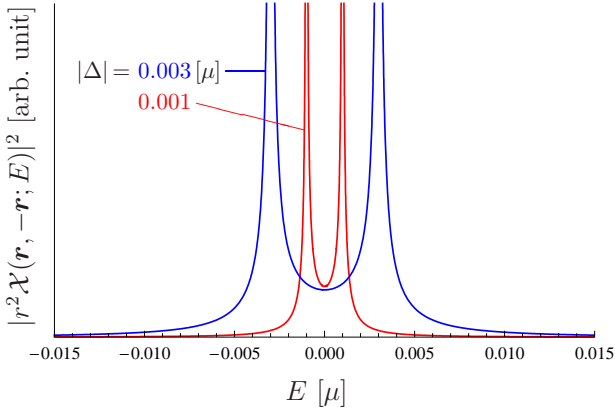}
\caption{(Color online) Spectrum of the Andreev emission for $\theta=\pi$ with (\ref{eqn:ChiThE})--(\ref{eqn:Chi0E}) and (\ref{eqn:Zth})--(\ref{eqn:Z0}).
The parameters are the same as those in Fig.\ \ref{fig:EnergySpectrum}.}
\label{fig:AndreevSpectrum}
\end{figure}
A remarkable feature of the spectrum $P(E)$ shown in Fig.\ \ref{fig:EnergySpectrum} is the presence of the gap for the superconducting emitter: no emission from the quasiparticles in the range $-|\Delta|<E<|\Delta|$.
In addition, the spectrum diverges at the edges of the gap, reflecting the density of states of the quasiparticles in the superconductor.

It should be noted, however, that we are looking at the two-point correlation functions at the lowest order in the tunneling Hamiltonian, i.e.\ up to $O(\lambda^2)$ [see (\ref{eqn:Gamma})--(\ref{eqn:Andreev})].
The correlation function $\gamma$ in (\ref{eqn:Gamma}) up to this order counts only the direct emission from the quasiparticle excitations, but there also exist emissions from the Cooper pairs.
Let us look at the spectrum of the other two-point correlation function, $\chi$, which describes the Andreev emission from the Cooper pairs.
Figure \ref{fig:AndreevSpectrum} shows the spectrum for $\bm{r}_1=-\bm{r}_2$, i.e., when the pair electrons are detected in opposite directions.
It exhibits nonzero spectrum even inside the gap of the quasiparticle spectrum.
This is the distinctive feature of the Andreev process.
If one collects higher-order processes, the single-particle spectrum $P(E)$ also accounts for the contribution of the Andreev emission, and one sees a spectrum also in the gap.\cite{ref:MauroSaro}

Suppose now that
\begin{equation}
\mu\gg E_C\,(\gg k_BT,\ \text{usually}),
\end{equation}
which is usually the case in actual experiments.
Then, only energies close to the Fermi level $|E|/\mu\ll1$ contribute to the correlation function $\gamma$.
As for the other spectra $\mathcal{X}_\text{th}$ and $\mathcal{X}_0$ in (\ref{eqn:ChiThE}) and (\ref{eqn:Chi0E}), the factor
\begin{equation}
\e^{ i[p(E)+p(-E)]r}
\simeq\e^{2i k_Fr[1-E^2/8\mu^2+O(E^4/\mu^4)]}
\label{eqn:FrequentFresnel}
\end{equation}
rapidly oscillates for $k_Fr\gg1$ and its stationary phase at the Fermi level $E=0$ provides the most significant contribution.
The relevant energy range in the saddle-point approximation is
\begin{equation}
|E|/\mu\lesssim\sqrt{2/k_Fr}\ll1.
\end{equation}
Therefore, only energies close to the Fermi level $|E|/\mu\ll1$ are relevant to the correlation functions.
In this regime, $Z(E)$, $Z_\text{th}(E)$, and $\delta Z_0(E)$ in (\ref{eqn:Zs}) are approximated by 
\begin{widetext}
\begin{subequations}
\begin{align}
&Z(E)
\simeq
\frac{1}{\cos(\theta/2)}
\,\Bigl(
\e^{-4k_F^2w^2\sin^2(\theta/4)(
1+E/\mu-|\Delta|^2/2\mu E
)}
-\e^{-4k_F^2w^2\cos^2(\theta/4)(
1+E/\mu-|\Delta|^2/2\mu E
)}
\Bigr),
\label{eqn:Zapp}
\displaybreak[0]\\
&Z_\text{th}(E)\e^{ i[p(E)+p(-E)]r}
\simeq\frac{\e^{2i k_Fr}}{\sin(\theta/2)}
\Bigl(
\e^{-4k_F^2w^2\sin^2[(\pi-\theta)/4]}
\e^{(w^2/\pi^2\xi^2)\sin(\theta/2)}\e^{- i E^2/2\kappa_+}
\nonumber\\[-2mm]
&\hspace*{55mm}
{}-\e^{-4k_F^2w^2\cos^2[(\pi-\theta)/4]}
\e^{-(w^2/\pi^2\xi^2)\sin(\theta/2)}
\e^{- i E^2/2\kappa_-}
\Bigr),
\label{eqn:Zthapp}
\displaybreak[0]\\
&\delta Z_0(E)\e^{ i[p(E)+p(-E)]r}
\simeq
\frac{ i k_Fw\Lambda}{\sqrt{\pi}}\frac{\sqrt{E^2-|\Delta|^2}}{\mu}
\frac{\e^{2i k_Fr}}{\sin(\theta/2)}
\e^{-k_F^2w^2\cos^2(\theta/2)}
\e^{- i E^2/2\kappa_0},
\label{eqn:Z0app}
\end{align}
\end{subequations}
\end{widetext}
where
\begin{subequations}
\label{eqn:KappaNu}
\begin{gather}
\kappa_\pm^{-1}
=\frac{k_F^2}{\mu^2}(r/2k_F\mp i\nu w^2),
\quad
\nu
=\sin\frac{\theta}{2}
-\frac{1}{2}\cos\frac{\theta}{2}\cot\frac{\theta}{2},
\displaybreak[0]
\label{eqn:Kappapm}
\\
\kappa_0^{-1}
=\frac{k_F^2}{2\mu^2}(r/k_F+ iw^2\cos\theta),
\end{gather}
\begin{multline}
\Lambda
=\frac{2}{\sqrt{\pi}}\left[
\Xi'\!\left(
2k_Fw\sin^2\!\frac{\pi-\theta}{4}
\right)
\right.
\\
\left.{}-\Xi'\!\left(
2k_Fw\cos^2\!\frac{\pi-\theta}{4}
\right)
\right].
\label{eqn:Lambda}
\end{multline}
\end{subequations}
Additional conditions have been assumed:
\begin{equation}
1/2k_Fr\sin^2(\theta/2)\ll1,\quad
\sqrt{k_Fw^2/2r}\ll1
\end{equation}
for (\ref{eqn:Zthapp}) and (\ref{eqn:Z0app}). 
Note that Eq.\ (\ref{eqn:Zapp}) is still exact for a normal emitter $\Delta=0$.

\subsection{Analytical Formulas for the Correlation Functions}
\label{sec:AnalyticalFormulas}
Several concise and useful analytical formulas are now available.
In the following, let
$\tau=t_1-t_2$.

\subsubsection{Normal emitter $\Delta=0$ at $T=0$}
In this case, $\chi$ vanishes, while the inverse Fourier transform of (\ref{eqn:GammaE}) with (\ref{eqn:Z}) [or equivalently, with (\ref{eqn:Zapp})] yields 
\begin{align}
\gamma(1;2)
&=\frac{A}{\cos(\theta/2)}
\,\biggl(
\frac{\e^{-4k_F^2w^2\sin^2(\theta/4)}
-\e^{-\mu(E_C^{-1}+ i\tau)}}{
E_C^{-1}+ i\tau-4(k_F^2w^2/\mu)\sin^2(\theta/4)
}
\nonumber
\\
&\quad\qquad\qquad
{}-\frac{\e^{-4k_F^2w^2\cos^2(\theta/4)}
-\e^{-\mu(E_C^{-1}+ i\tau)}}{
E_C^{-1}+ i\tau-4(k_F^2w^2/\mu)\cos^2(\theta/4)
}\biggr).
\label{eqn:GammaNormalZeroTemp}
\end{align}

\subsubsection{Normal emitter $\Delta=0$ at $T>0$}
Even at finite temperature $T>0$, if
\begin{equation}
1/k_BT>1/E_C-4k_F^2w^2/\mu>0
\quad\text{and}\quad
\e^{-\mu/E_C}\ll1
\label{eqn:ConditionFiniteTemp}
\end{equation}
are satisfied, 
an analytical formula is available: the lower end of 
the integration range in energy can be extended to $-\mu\to-\infty$ and one gets
\begin{widetext}
\begin{equation}
\gamma(1;2)
\simeq\frac{A\pi k_BT}{\cos(\theta/2)}\,
\left(
\frac{\e^{-4k_F^2w^2\sin^2(\theta/4)}}{\sin\{
\pi k_BT[E_C^{-1}+ i\tau-4(k_F^2w^2/\mu)\sin^2(\theta/4)]
\}}
-\frac{\e^{-4k_F^2w^2\cos^2(\theta/4)}}{\sin\{
\pi k_BT[E_C^{-1}+ i\tau-4(k_F^2w^2/\mu)\cos^2(\theta/4)]
\}}
\right)
\label{eqn:GammaNormalTemp}
\end{equation}
\end{widetext}
by making use of the formula (\ref{eqn:IntForm}).

\subsubsection{For superconducting emitter $|\Delta|>0$ at $T=0$}
The inverse Fourier transform of (\ref{eqn:GammaE}) with (\ref{eqn:Zapp}) yields
\begin{align}
\gamma(1;2)
\simeq{}&\frac{A}{\cos(\theta/2)}
|\Delta|K_1\bm{(}|\Delta|(E_C^{-1}+ i\tau)\bm{)}
\nonumber
\\
&{}\times
\Bigl(
\e^{-4k_F^2w^2\sin^2(\theta/4)}
-\e^{-4k_F^2w^2\cos^2(\theta/4)}
\Bigr)
\label{eqn:GammaSuper}
\end{align}
by neglecting $O(E_C/\mu)$ contributions, and that of (\ref{eqn:Chi0E}) with (\ref{eqn:Zthapp}) and (\ref{eqn:Z0app}) gives
\begin{equation}
\chi_0(1;2)
=\pi A
\int_{-\infty}^\infty\d\tau'\,
F(\tau-\tau')G(\bm{r}_1,\bm{r}_2;\tau')
+\delta\chi_0(1;2)
\label{eqn:XiConv}
\end{equation}
with
\begin{equation}
F(\tau)
=\int_{-\infty}^\infty\frac{\d E}{2\pi}
\frac{\Delta}{\sqrt{E^2-|\Delta|^2}}
\e^{ i E\tau}
=\frac{\Delta}{2i}H_0^{(2)}(|\Delta\tau|),
\label{eqn:Fana}
\end{equation}
\begin{align}
&G(\bm{r}_1,\bm{r}_2;\tau)
\nonumber\displaybreak[0]\\
&\quad
=\int_{-\infty}^\infty\frac{\d E}{2\pi}\,
Z_\text{th}(E)\e^{ i[p(E)+p(-E)]r}\e^{ i E\tau}
\nonumber
\displaybreak[0]
\\
&\quad
\simeq\frac{\e^{2i k_Fr}}{\sin(\theta/2)}
\e^{-4k_F^2w^2\sin^2[(\pi-\theta)/4]}
\e^{(w^2/\pi^2\xi^2)\sin(\theta/2)}
\nonumber
\\[-1mm]
&\hspace*{57mm}
{}\times\sqrt{\frac{\kappa_+}{2\pi i }}\e^{ i\kappa_+\tau^2/2},
\label{eqn:AppG}
\end{align}
and
\begin{align}
&\delta\chi_0(1;2)
\nonumber\\
&\ \ %
=\frac{A}{2}\int_{-\infty}^\infty\d E\,
\frac{\Delta}{\sqrt{|E|^2-|\Delta|^2}}
\,\delta Z_0(E)\e^{ i[p(E)+p(-E)]r}\e^{ i E\tau}
\nonumber
\displaybreak[0]
\\
&\ \ %
\simeq\frac{\sqrt{\pi}\, i  A\Delta}{\sin(\theta/2)}
\frac{k_Fw\Lambda}{\mu}
\e^{2i k_Fr}
\e^{-k_F^2w^2\cos^2(\theta/2)}
\sqrt{\frac{\kappa_0}{2\pi i }}\e^{ i\kappa_0\tau^2/2}.
\label{eqn:Gana}
\end{align}
For the coincident detections with $\tau=t_1-t_2=0$, another expression is convenient:
\begin{widetext}
\begin{multline}
\chi_0(1;2)
\simeq\frac{\pi A\Delta\e^{2i k_Fr}}{4i \sin(\theta/2)}
\,\Biggl(
\e^{-4k_F^2w^2\sin^2[(\pi-\theta)/4]}
\e^{- i r/2\pi^2k_F\xi^2}
\e^{-(w^2/\pi^2\xi^2)[\nu-\sin(\theta/2)]}
H_0^{(2)}\bm{(}
( i\nu w^2-r/2k_F)/\pi^2\xi^2
\bm{)}
\\
-\frac{4\Lambda\e^{-k_F^2w^2\cos^2(\theta/2)}}{\pi\sqrt{ i r/k_Fw^2}}
\Biggr).
\label{eqn:Chi0Coinc}
\end{multline}
In (\ref{eqn:AppG}) and (\ref{eqn:Chi0Coinc}), the second contribution in (\ref{eqn:Zthapp}) has been omitted by assuming $\e^{-2k_F^2w^2}\ll1$.
As for the Bessel functions $K_\nu(z)$ and $H_\nu^{(2)}(z)$, see Appendix \ref{app:IntForm}\@.
Note that $\Lambda$ defined in (\ref{eqn:Lambda}) is almost constant $\Lambda\simeq1$ for $w\gtrsim \lambda_F$ 
\end{widetext}
and $\theta\simeq\pi$.

An analytical formula is not available for the superconducting emitter $|\Delta|>0$ at a finite temperature $T>0$.
In the present analysis, the effects of temperature are taken into account only though the Fermi distribution function $f(E)$, which controls the high-energy tails in the spectra.
They are usually small, with $E_C\gg k_BT$ like in Fig.\ \ref{fig:EnergySpectrum}, even at room temperature.
For instance, $\pi k_BT/\sin(\pi k_BT/E_C)\simeq E_C$ in formula (\ref{eqn:GammaNormalTemp}) (almost no temperature effect), and the other analytical formulas for $T=0$ agree well with the numerical estimations for the finite-temperature case with the value chosen in Fig.\ \ref{fig:EnergySpectrum}.
It therefore suffices in the following discussion, to use the formulas for $T=0$.
Physically, the gap parameter $\Delta$ is a function of temperature $T$, so that
temperature effects appear through the gap parameter $|\Delta|$.

\section{Antibunching and Bunching}
\label{sec:BunchingAntibunching}
\begin{figure}[t]
\includegraphics[width=0.45\textwidth]{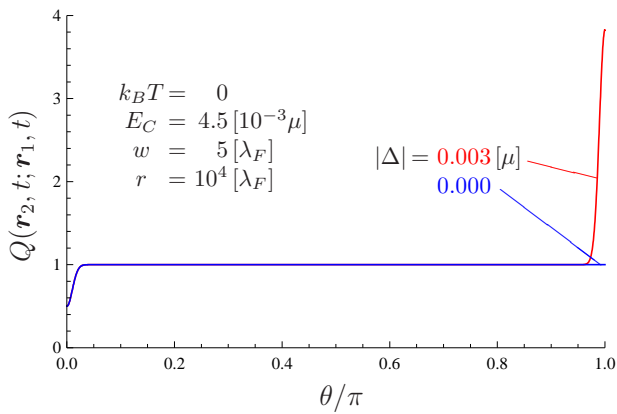}
\caption{(Color online) Normalized coincidence $Q$ vs $\theta$ for $t_1=t_2$, for normal and superconducting emitters.}
\label{fig:CorrelationAngle}
\end{figure}

We are now ready to discuss the correlation between the electrons emitted from a superconductor.
The normalized coincidence $Q$ is shown in Fig.\ \ref{fig:CorrelationAngle} as a function of $\theta$ at $t_1=t_2=t$, for normal and superconducting emitters.

\subsection{Antibunching at $\theta\sim0$}
The dip around $\theta\sim0$ is due to the antibunching of the electrons: 
there are fewer chances to detect two electrons at the same time than expected from the counting rate of the independent detections of single electrons.
The dip is $Q\simeq0.5$, irrespectively of the parameters.
This is clear from formula (\ref{eqn:NormalizedCorrelationST}) for $Q$.
Figure \ref{fig:CorrelationAngle} shows that the superconductivity of the emitter does not affect the antibunching and $\chi\simeq0$ for $\theta\sim0$.
Hence, formula (\ref{eqn:NormalizedCorrelationST}) yields $Q\simeq0.5$ for $\theta=0$ and $t_1=t_2$, irrespectively of the details of the correlation function $\gamma$.
Equation (\ref{eqn:TripSingPair}) shows that three antibunching contributions plus one bunching yields $1-3/4+1/4=0.5$.\cite{ref:Antibunching-Neutron,ref:LateralEffects}

\begin{figure}[t]
\includegraphics[width=0.45\textwidth]{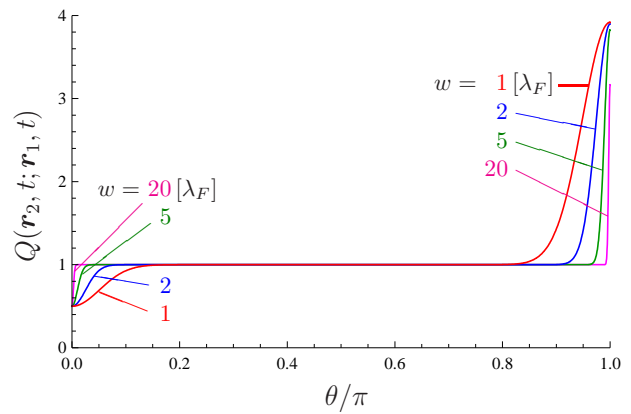}
\caption{(Color online) Normalized coincidence $Q$ vs $\theta$ for $t_1=t_2$, for a superconducting emitter with different values of $w$.
The other parameters are the same as those in Fig.\ \ref{fig:CorrelationAngle} with $|\Delta|=0.003\,[\mu]$.}
\label{fig:CorrelationAngleW}
\end{figure}

The analytical formulas for $\gamma$ in Sec.\ \ref{sec:AnalyticalFormulas} show that the width of the dip is governed by the factor $\e^{-8k_F^2w^2\sin^2(\theta/4)}$ and is controlled by the size of the emitting region, $w$.
The smaller the size of the emitting region, the longer the lateral coherence length of the emitted electrons, and the wider the antibunching dip.\cite{ref:LateralEffects}
See Fig.\ \ref{fig:CorrelationAngleW}, where the effect of $w$ is scrutinized.

\subsection{Bunching at $\theta\sim\pi$}
\label{sec:Bunching}
The effect of the superconducting emitter manifests itself when the electrons are detected in opposite directions, $\theta\sim\pi$.
Since a Cooper pair is formed by two electrons with opposite momenta $\bm{k}$ and $-\bm{k}$, they are emitted in opposite directions and exhibit strong correlation at $\theta\sim\pi$.
The spins of the Cooper pair are in the singlet state, and the corresponding symmetric wave function in momentum space gives rise to bunching, namely a positive correlation, as already discussed in Sec.\ \ref{sec:Structure}.

The analytical formulas for $\chi$ in Sec.\ \ref{sec:AnalyticalFormulas} show that the width of the bunching peak is characterized by the factor $\e^{-8k_F^2w^2\sin^2[(\pi-\theta)/4]}$, and the lateral coherence length is the same as that for the antibunching.
The Cooper pair does not affect the lateral coherence.

The height of the bunching peak $\delta Q_\text{peak}=Q(\bm{r},t,-\bm{r},t)-1$ is, on the other hand, given by
\begin{align}
\delta Q_\text{peak}
&\simeq\frac{|\chi(\bm{r},t;-\bm{r},t)|^2}{2|\gamma(\bm{r},t;\bm{r},t)|^2}
\nonumber
\displaybreak[0]
\\
&
\simeq\frac{\pi^2}{32K_1^2(|\Delta|/E_C)}
\,\Biggl|
H_0^{(2)}\bm{(}
( iw^2-r/2k_F)/\pi^2\xi^2
\bm{)}
\nonumber\\[-1mm]
&\qquad\qquad\qquad\qquad\qquad\qquad
{}
-\frac{
4\Lambda\e^{ i r/2\pi^2k_F\xi^2}
}{\pi\sqrt{ i r/k_Fw^2}}
\Biggr|^2,
\label{eqn:Peak}
\end{align}
the ratio of (\ref{eqn:Chi0Coinc}) to (\ref{eqn:GammaSuper}).
This is the ratio of the contribution of the pair emission to that of the single-particle emission, the background.
The main contribution to the former is provided by the term represented by the Hankel function $H_0^{(2)}$ when $w\ll\xi$: it is greater than the other by a factor of order $\xi/w$ [see for instance the asymptotic form of the Hankel function in (\ref{eqn:BesselHCompAsymp})].
This pair emission is controlled by the parameters $r/2k_F\xi^2$ and $w^2/\xi^2$, both measured against Pippard's length $\xi$, while the single-particle emission by $|\Delta|/E_C$.
In particular, the emission from a smaller emitting region results in a stronger correlation.
See Fig.\ \ref{fig:CorrelationAngleW}, which clarifies how the size of the emitting region, $w$, affects the bunching peak.

\subsection{Correlation in Time}
\begin{figure*}
\begin{tabular}{ll}
(a)&(b)\\[-3.6mm]
\includegraphics[width=0.475\textwidth]{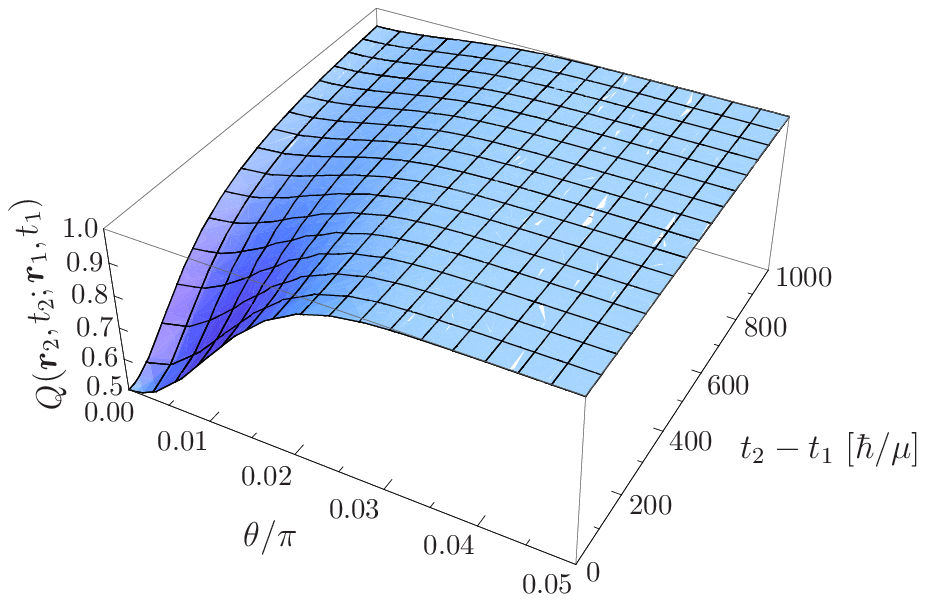}&
\includegraphics[width=0.475\textwidth]{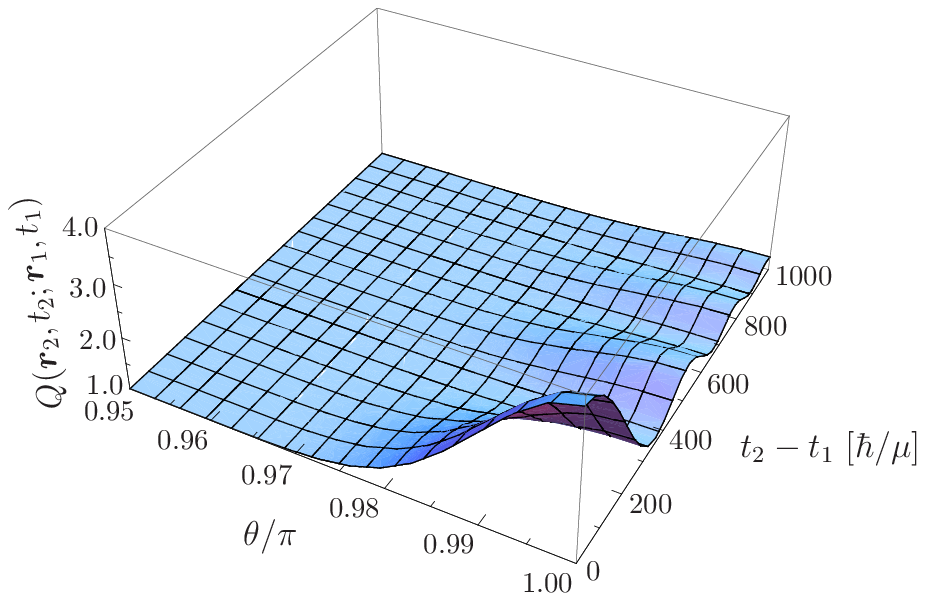}
\end{tabular}
\caption{(Color online) Normalized coincidence $Q$ as a function of $\theta$ and delay time $\tau=t_2-t_1$.  Two different regions of $\theta$ are closed up in (a) and (b).  The parameters are the same as those in Fig.\ \ref{fig:CorrelationAngle} with $|\Delta|=0.003\,[\mu]$.}
\label{fig:CorrelationAngleTime}
\end{figure*}

\begin{figure}
\includegraphics[width=0.45\textwidth]{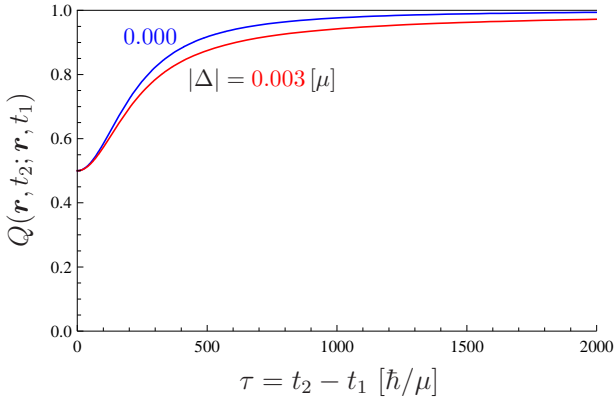}
\caption{(Color online) Normalized coincidence $Q$ vs delay time $\tau=t_2-t_1$ for $\theta=0$, for normal and superconducting emitters.
The parameters are the same as those in Fig.\ \ref{fig:CorrelationAngle}.}
\label{fig:CorrelationTime0}
\end{figure}

Let us now look at correlations in time (Fig.\ \ref{fig:CorrelationAngleTime}).
Although superconductivity does not affect the depth of the antibunching dip and the lateral coherence as is clear from Fig.\ \ref{fig:CorrelationAngle}, it influences the correlation function in time even at $\theta\sim0$ [Fig.\ \ref{fig:CorrelationAngleTime}(a)].
The analytical formulas (\ref{eqn:GammaNormalZeroTemp}) and (\ref{eqn:GammaNormalTemp}) show that the correlation function $Q$ at $\theta=0$ decays as a function of the delay time $\tau=t_2-t_1$ like a Lorentzian $\sim(1+E_C^2\tau^2)^{-1}$ for a normal emitter, while formula (\ref{eqn:GammaSuper}) reveals that the decay for the superconducting emitter is like $\sim(\Delta\tau)^{-1}$ with a different power tail [see (\ref{eqn:BesselKAsymp}) for the asymptotic behavior of $K_\nu(z)$].
Both cases are compared in Fig.\ \ref{fig:CorrelationTime0}.
This prolonged tail is due to the divergence in the spectrum at the edge of the gap (Fig.\ \ref{fig:EnergySpectrum}) and is an effect of the superconductor.
\begin{figure}
\begin{tabular}{l}
(a)\\[-3.9mm]
\includegraphics[width=0.45\textwidth]{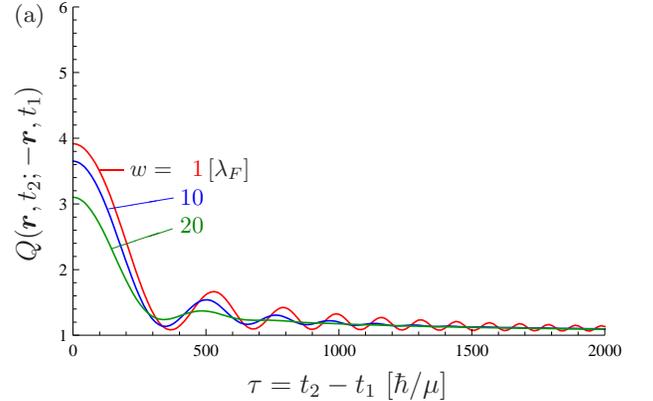}\\[2mm]
(b)\\[-3.9mm]
\includegraphics[width=0.45\textwidth]{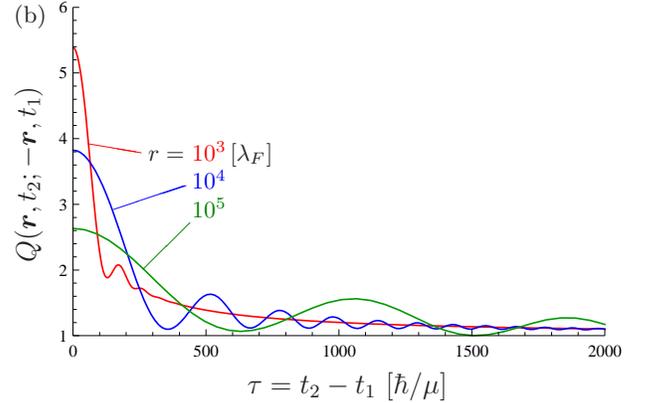}
\end{tabular}
\caption{(Color online) Normalized coincidence $Q$ vs delay time $\tau=t_2-t_1$ for $\theta=\pi$, (a) with different values of $w$ and (b) with different values of $r$.
Other parameters are the same as those in Fig.\ \ref{fig:CorrelationAngle} with $|\Delta|=0.003\,[\mu]$.}
\label{fig:CorrelationFourierWR}
\end{figure}

At $\theta\sim\pi$, where the superconductivity of the emitter prominently manifests itself, the correlation function exhibits oscillatory decay in time [Fig.\ \ref{fig:CorrelationAngleTime}(b)].
This is governed by the convolution in (\ref{eqn:XiConv}): the damped oscillation $\e^{ i\kappa_+\tau^2/2}$ in (\ref{eqn:AppG}) is convoluted with the slow decay of $|H_0^{(2)}(|\Delta\tau|)|\sim|\Delta\tau|^{-1/2}$ in (\ref{eqn:Fana}) [see (\ref{eqn:BesselHRealAsymp}) for the asymptotic behavior of $H_\nu^{(2)}(x)$].
The time scales of the damped oscillation are given by (\ref{eqn:Kappapm}),
\begin{equation}
 i\kappa_+/2
=\mu^2\frac{ i/k_Fr-2\nu w^2/r^2}{1+(2\nu k_Fw^2/r)^2}.
\end{equation}
See Fig.\ \ref{fig:CorrelationFourierWR}, which demonstrates how the parameters $w$ and $r$ control the damped oscillation.

\section{Entanglement}
\label{sec:Entanglement}
We now turn our attention to the state corresponding to the antibunching and bunching correlations.
As already mentioned in Sec.\ \ref{sec:Structure}, a pair of electrons emitted through the Andreev process forms a singlet spin state, reflecting the singlet state of the Cooper pair.
It is interesting to explore the possibility that a superconductor be a source of entanglement.\cite{ref:Loss,ref:Blatter,ref:SauretFeinbergMartin,ref:Buettiker,ref:Prada,ref:FaoroTaddeiFazio,ref:review-Beenakker,ref:review-Burkard}

Let the two angles $(\eta,\zeta)$ specify the orientation of the spin of an electron, the polar and azimuthal angles with respect to a certain quantization axis of the spin.
The probability that an electron is found at $(\bm{r}_1,t_1)$ with its spin oriented in the direction $(\eta_1,\zeta_1)$ and another at $(\bm{r}_2,t_2)$ ($t_2\ge t_1$) with $(\eta_2,\zeta_2)$ is proportional to
\begin{equation}
\rho_{\eta_2\zeta_2,\eta_1\zeta_1}(2;1)
=\langle
\psi_{\eta_1\zeta_1}^\dag(1)
\psi_{\eta_2\zeta_2}^\dag(2)
\psi_{\eta_2\zeta_2}(2)
\psi_{\eta_1\zeta_1}(1)
\rangle,
\end{equation}
where
\begin{equation}
\psi_{\eta\zeta}(\bm{r},t)
=\psi_\uparrow(\bm{r},t)\cos\frac{\eta}{2}
+\psi_\downarrow(\bm{r},t)\e^{- i\zeta}\sin\frac{\eta}{2}
\end{equation}
is the field that annihilates an electron with its spin oriented along $(\eta,\zeta)$.
This two-particle distribution function is cast into the following form through Wick's theorem:
\begin{align}
&\rho_{\eta_2\zeta_2,\eta_1\zeta_1}(2;1)
\nonumber
\\
&\quad
=\gamma(2;2)\gamma(1;1)
\nonumber
\\
&\quad\quad
{}-|\gamma(2;1)|^2
\left|
\cos\frac{\eta_1}{2}\cos\frac{\eta_2}{2}
+\e^{ i(\zeta_1-\zeta_2)}
\sin\frac{\eta_1}{2}\sin\frac{\eta_2}{2}
\right|^2
\nonumber
\displaybreak[0]
\\
&\quad\quad
{}+|\chi(2;1)|^2
\left|
\e^{ i\zeta_1}\sin\frac{\eta_1}{2}\cos\frac{\eta_2}{2}
-\e^{ i\zeta_2}\cos\frac{\eta_1}{2}\sin\frac{\eta_2}{2}
\right|^2.
\label{eqn:TwoDistriSpin}
\end{align}
All the matrix elements of the density operator $\varrho$ of the spins of a pair of electrons found at $(\bm{r}_1,t_1)$ and $(\bm{r}_2,t_2)$ ($t_2\ge t_1$) are reconstructed by combining $\rho_{\eta_2\zeta_2,\eta_1\zeta_1}(2;1)$ with appropriate sets of the angles to yield 
\begin{align}
\varrho\propto{}&
\Bigl(
\gamma(2;2)\gamma(1;1)
-|\gamma(2;1)|^2
\Bigr)\,
\openone\nonumber\displaybreak[0]\\
&{}+2\,\Bigl(
|\gamma(2;1)|^2+|\chi(2;1)|^2
\Bigr)\,\ket{\Psi^-}\bra{\Psi^-},
\label{eqn:ExtractedState}
\end{align}
where $\ket{\Psi^-}=(\ket{\uparrow\downarrow}-\ket{\downarrow\uparrow})/\sqrt{2}$ is the singlet state.
Note that the normalization factor is given by the two-particle distribution $\rho^{(2)}(2;1)$ in (\ref{eqn:TwoDistri}).

The second term of (\ref{eqn:ExtractedState}) is the singlet state, while the first one is the background that blurs the entanglement.
At $\theta=0$ (when the two detectors are at the same place), the coincident detections with $t_1=t_2$ eliminate the background and a pure singlet state is extracted.
This is because the antisymmetric wave function (the triplet contribution) does not trigger two detectors located at the same point, where the antisymmetric wave function is vanishing. 
In this way, the triplet components are ruled out by the coincident detections and the singlet component is extracted.
The entanglement for $\theta\sim0$ is just due to the symmetry of the wave function and is not supplied by the superconductor.

A remarkable feature of this entanglement is that it requires no interaction between the electrons.
See Ref.\ \onlinecite{ref:BosonEntNonInt} for a similar mechanism to extract entanglement, making use of the Bose/Fermi symmetry of two-particle wave function.
For electronic systems, the interaction-free sources of electron-hole entanglement in solid are proposed in Ref.\ \onlinecite{ref:EntEleHole} (although the mechanism is different from the present one).

Let us look at the concurrence \cite{ref:Concurrence} of the state $\varrho$ in (\ref{eqn:ExtractedState}) extracted by the two detectors.
It is given by 
\begin{equation}
C(\varrho)=\max\!\left(
0,\frac{
|\chi(2;1)|^2+2|\gamma(2;1)|^2-\gamma(2;2)\gamma(1;1)
}{
2\gamma(2;2)\gamma(1;1)-|\gamma(2;1)|^2+|\chi(2;1)|^2
}
\right).
\label{eqn:Concurrence}
\end{equation}
This actually gives $C(\varrho)=1$ at $\theta=0$.
It is possible to show that the concurrence (\ref{eqn:Concurrence}) is nonzero, $C(\varrho)>0$, if
\begin{equation}
Q<3/4,\quad 3/2<Q
\end{equation}
[note that $\chi$ is essentially vanishing for $\theta\sim0$ and $\gamma$ is essentially vanishing for $\theta\sim\pi$].

The electron pair can even violate Bell's inequality.\cite{ref:Kawabata}
A nonlocal correlation between the two electrons is disclosed if the Clauser-Horne-Shimony-Holt (CHSH) inequality \cite{ref:CHSH}
\begin{equation}
|S|\le2
\label{eqn:CHSH}
\end{equation}
is violated with a certain set of angles $\eta_{1,2}$ and $\eta_{1,2}'$, where
\begin{equation}
S=E(\eta_1,\eta_2)+E(\eta_1,\eta_2')-E(\eta_1',\eta_2)+E(\eta_1',\eta_2')
\end{equation}
and
\begin{equation}
E(\eta_1,\eta_2)
=\frac{\rho_{\eta_1,\eta_2}+\rho_{\eta_1+\pi,\eta_2+\pi}-\rho_{\eta_1,\eta_2+\pi}-\rho_{\eta_1+\pi,\eta_2}}{\rho_{\eta_1,\eta_2}+\rho_{\eta_1+\pi,\eta_2+\pi}+\rho_{\eta_1,\eta_2+\pi}+\rho_{\eta_1+\pi,\eta_2}}
\end{equation}
with $\rho_{\eta_1,\eta_2}=\rho_{\eta_1\zeta,\eta_2\zeta}$.
In the present case, the two-particle distribution function $\rho_{\eta_1\zeta_1,\eta_2\zeta_2}$ in (\ref{eqn:TwoDistriSpin}) yields
\begin{equation}
E(\eta_1,\eta_2)
=-D\cos(\eta_1-\eta_2)
\label{eqn:CorrelationForBell}
\end{equation}
with a factor
\begin{equation}
D=\frac{
|\gamma(2;1)|^2
+|\chi(2;1)|^2
}{
2\gamma(2;2)\gamma(1;1)
-|\gamma(2;1)|^2
+|\chi(2;1)|^2
}.
\label{eqn:PrefactorForBell}
\end{equation}
Notice here that $E(\eta_1,\eta_2)=-\cos(\eta_1-\eta_2)$ represents the singlet spin state, and it maximally violates the CHSH inequality (\ref{eqn:CHSH}) with e.g.\ $\eta_1=0$, $\eta_2=-\pi/4$, $\eta_1'=\pi/2$, and $\eta_2'=\pi/4$, yielding $S=-2\sqrt{2}$.
The prefactor $D\,(\le1)$ in (\ref{eqn:PrefactorForBell}) shrinks the magnitude of $S$ and reduces the chance of the violation of the CHSH inequality (\ref{eqn:CHSH}).
The CHSH inequality (\ref{eqn:CHSH}) can thus be violated only when $D>1/\sqrt{2}$.
This is accomplished when
\begin{equation}
Q<\sqrt{2}/(\sqrt{2}+1),\ \sqrt{2}/(\sqrt{2}-1)<Q.
\end{equation}
In particular, at $\theta=0$ (when the two detectors are at the same place), $D=1$ and the coincident detections extract the singlet state, in accordance with the above argument.

The electron pair for $\theta\sim\pi$, corresponding to the bunching peak, can also be entangled and can violate Bell's inequality, provided the bunching is strong enough.
In this case, the paired electrons are emitted in opposite directions, and this is a more suitable situation for discussing nonlocality.
This channel is also preferable as a source of entanglement, since one need not argue how to separate the entangled electrons.\cite{note:Separate}
This is a remarkable contrast to the transport setups, in which it is generally hard to split the entangled pair.

\begin{figure*}
\begin{tabular}{r@{\qquad\qquad}r}
(a)&(b)\\
\includegraphics[height=0.25\textwidth]{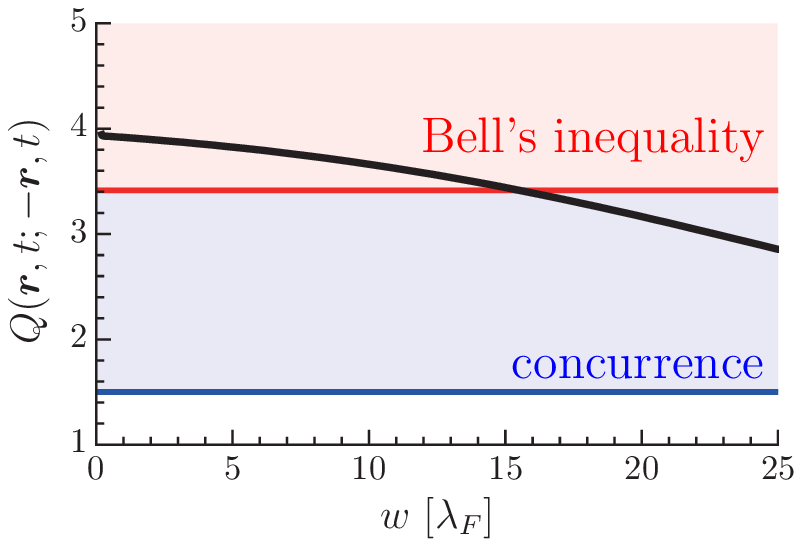}&
\includegraphics[height=0.25\textwidth]{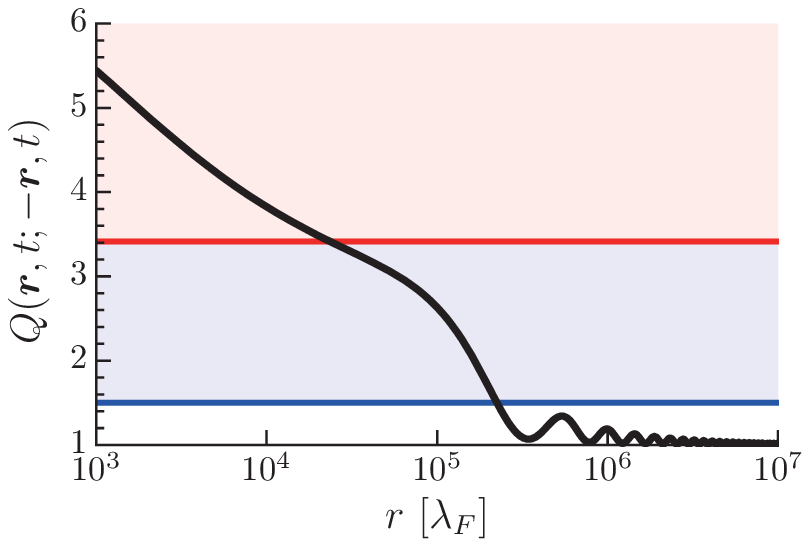}\\[3mm]
(c)&(d)\\
\includegraphics[height=0.25\textwidth]{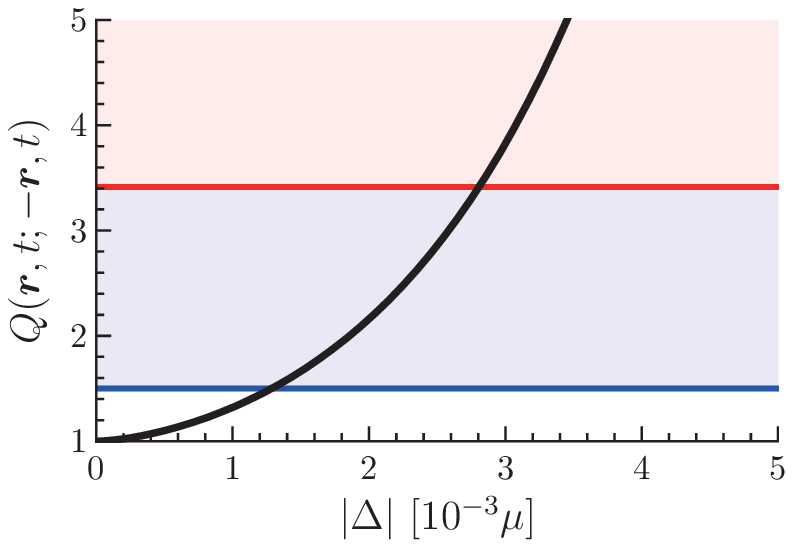}&
\includegraphics[height=0.25\textwidth]{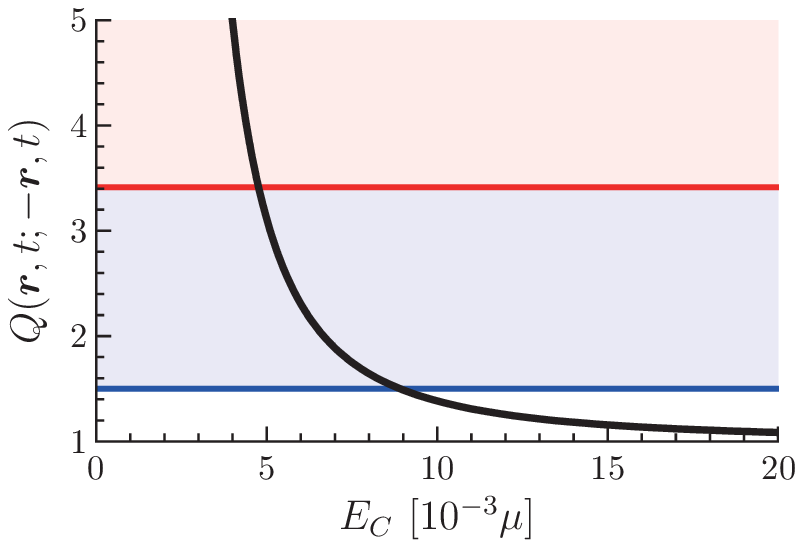}\\[3mm]
(e)&(f)\\
\includegraphics[height=0.25\textwidth]{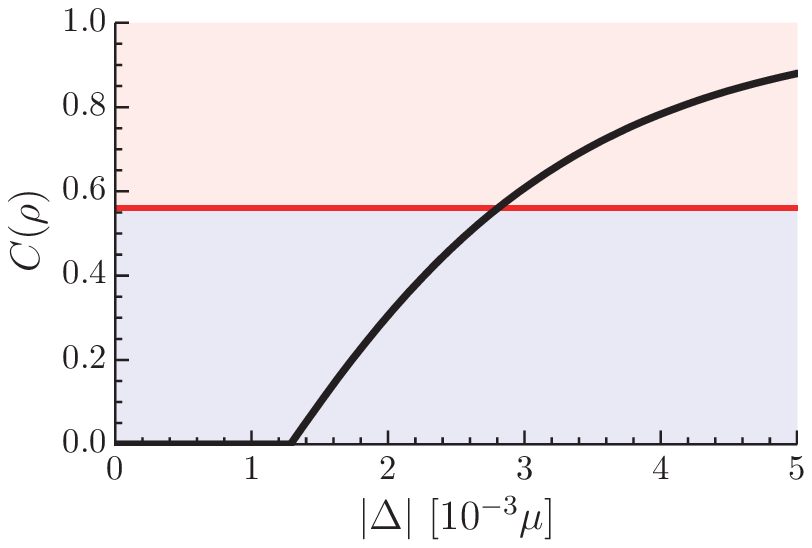}&
\includegraphics[height=0.25\textwidth]{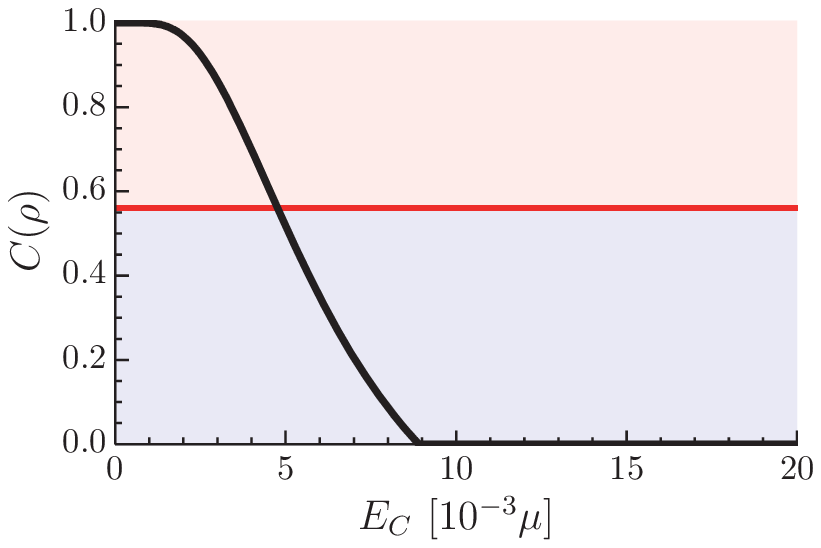}
\end{tabular}
\caption{(Color online) (a)--(d) The peak of the normalized coincidence $Q$ at $\theta=\pi$ and $t_1=t_2$, based on the analytical formula (\ref{eqn:Peak}).
The line at $Q=1.5$ indicates the threshold for nonzero concurrence $C(\varrho)>0$ and the one at $Q=\sqrt{2}/(\sqrt{2}-1)\simeq3.41$ is that for the violation of Bell's inequality.
(e)--(f) The concurrence $C(\varrho)$ corresponding to (c)--(d).
The line at $C(\varrho)=(3\sqrt{2}-2)/4\simeq0.56$ corrseponds to the threshold for the violation of Bell's inequality.
The parameters are the same as those in Fig.\ \ref{fig:CorrelationAngle} with $|\Delta|=0.003\,[\mu]$.
The correlation length of the superconductor is $\xi\simeq 33.8\,[\lambda_F]$ for this gap parameter.
}
\label{fig:Peak}
\end{figure*}
The bunching peak at $\theta=\pi$ is given by formula (\ref{eqn:Peak}).
There are essentially four independent parameters that control the bunching peak, and hence the entanglement, i.e.\ $w$, $r$, $\Delta$, and $E_C$.
See Fig.\ \ref{fig:Peak}, where the peak and the concurrence are plotted as functions of these parameters.

It is obvious that the gap parameter $\Delta$ plays a significant role for entanglement by enhancing the contribution of the Cooper pairs.
By increasing $|\Delta|$, the gap becomes wider, and the Andreev emission becomes dominant.
The enhancement of the entanglement with a wider gap is an evidence for the importance of the Andreev emission to entanglement.
The parameter $E_C$ also works like a filter for the Andreev emission.
By decreasing $E_C$, single-particle emissions from the quasiparticle excitations are suppressed, and in this way, the background is reduced.
As a result, the pair emissions become dominant and the entanglement is enhanced.

It is interesting to observe that the size of the emitting region, $w$, also affects the entanglement.
The electron pairs bunch and are entangled only when they come from a small region.
This is because the emission from a small region ensures that a pair comes from a common Cooper pair in the emitter.
(If the pair electrons come from different Cooper pairs, they are not correlated and do not bunch.)
That is why the ratio $w/\xi$, between the size of the emitting region $w$ and the extension of the Cooper pair $\xi$, enters formula (\ref{eqn:Peak}) for the peak of the bunching and rules the entanglement of the emitted pair.

This is essentially equivalent to the effect due to the crossed Andreev reflection, which has been extensively studied in transport problems,\cite{ref:Loss,ref:Blatter,ref:SauretFeinbergMartin,ref:Buettiker,ref:Prada,ref:FaoroTaddeiFazio,ref:review-Burkard,ref:PositiveCorr,ref:CrossedAndreevTh,ref:CrossedAndreevExp} where an electron injected from one normal lead to a superconductor is reflected to the other normal lead as a hole, and vice versa.
It is shown that the crossed Andreev reflections are suppressed as the distance between the contacts of the normal leads to the superconductor is extended.\cite{ref:Loss,ref:SauretFeinbergMartin,ref:Prada,ref:CrossedAndreevTh,ref:CrossedAndreevExp}
It is instructive to rewrite the pair correlation function $\chi_0$ given in (\ref{eqn:Andreev}) in the following way.
At far places from the emitter, the first contribution in (\ref{eqn:Andreev}) survives and only energies close to the Fermi level contribute to $\chi_0$.
By putting $\varepsilon_{p_i}\simeq0$ in $\varepsilon_{p_i}+\omega_k-i0^+$ ($i=1,2$) in the denominators and by substituting the integral expression for the tunneling matrix element $T_{\bm{p}\bm{k}}$ given in (\ref{eqn:EmissionMatrix}), the pair correlation $\chi_0$ is cast in the form
\begin{align}
&\chi_0(\bm{r}_1,t_1;\bm{r}_2,t_2)
\nonumber\\
&\qquad
\simeq\lambda^2
\int d^3\bm{r}_1'\,d^3\bm{r}_2'\,
\Phi(\bm{r}_1-\bm{r}_1',t_1;\bm{r}_2-\bm{r}_2',t_2)
\nonumber\\
&\qquad\qquad\qquad\qquad\qquad\qquad{}\times
g(\bm{r}_1')
g(\bm{r}_2')
\mathcal{F}(\bm{r}_1'-\bm{r}_2')
\label{eqn:CA2W}
\end{align}
with
\begin{align}
\Phi(\bm{r}_1,t_1;\bm{r}_2,t_2)
={}&\int\frac{\d^3\bm{p}_1}{\sqrt{(2\pi)^3}}
\frac{\d^3\bm{p}_2}{\sqrt{(2\pi)^3}}\,
h(\bm{p}_1)
h(\bm{p}_2)
\nonumber\\
&\qquad\quad{}\times
\frac{\e^{i\bm{p}_1\cdot\bm{r}_1}\e^{i\bm{p}_2\cdot\bm{r}_2}}{\varepsilon_{p_1}+\varepsilon_{p_2}-i0^+}
\e^{- i\varepsilon_{p_1}(t_1-t_2)}
\end{align}
and
\begin{equation}
\mathcal{F}(\bm{r})
=\int\frac{\d^3\bm{k}}{(2\pi)^6}\,\frac{2u_kv_k}{\omega_k}e^{i\bm{k}\cdot\bm{r}}
\simeq e^{i\delta}\frac{mk_F}{(2\pi)^4}
\frac{\sin k_Fr}{k_Fr}
e^{-r/\pi\xi}
\label{eqn:CrossedAndreev}
\end{equation}
for $|\Delta|\ll\mu$.
The function $\mathcal{F}(\bm{r})$ in (\ref{eqn:CrossedAndreev}) is exactly the one often found in the arguments on the crossed Andreev reflections in the literature,\cite{ref:Loss,ref:SauretFeinbergMartin,ref:Prada,ref:review-Burkard,ref:CrossedAndreevTh,ref:CrossedAndreevExp} which describes the correlation between the electrons emitted at different points $\bm{r}_1'$ and $\bm{r}_2'$ in the superconductor and decays as the emitting points $\bm{r}_1'$ and $\bm{r}_2'$ are separated far away.
The emitted electrons are propagated by $\Phi(\bm{r}_1-\bm{r}_1',t_1;\bm{r}_2-\bm{r}_2',t_2)$ in vacuum, from $\bm{r}_1'$ to $\bm{r}_1$ and $\bm{r}_2'$ to $\bm{r}_2$, and such processes are integrated over the emitting points $\bm{r}_1'$ and $\bm{r}_2'$ in the source with a weight function $g(\bm{r})$, chosen to be Gaussian of size $w$ in (\ref{eqn:Setup}) in the present analysis.
This is the physical structure of $\chi_0$.
The formula (\ref{eqn:CA2W}) clarifies the connection between the effect of the size of the source, $w$, and the crossed Andreev effects.

The bunching peak decays as a function of $r$.
It decays like $\sim(r/k_F\xi^2)^{-1}$ with oscillations around it [see (\ref{eqn:Peak}) and the asymptotic behavior of $H_\nu^{(2)}( i z)$ in (\ref{eqn:BesselHCompAsymp})].
This oscillation originates from the divergences at the edges of the gap, $E=\pm|\Delta|$.
The reason why the bunching peak decays as a function of $r$ is the following.
The wave packets of the emitted electrons spread as they propagate.
Even if two electrons are detected at the same distance in opposite directions, this does not ensure that the two electrons originate from a common Cooper pair.
There is an ambiguity to the extent of the spreads of the wave packets.
The propagator in the free space explains that the uncertainty develops up to $\lesssim\sqrt{t/m}$ after an elapsed time $t$, where $t$ is translated into $r$ via $t\sim mr/k_F$.
This uncertainty should be smaller than $\xi$ for the two electrons to bunch.
This is the reason why the bunching peak (\ref{eqn:Peak}) decays for $r\gg k_F\xi^2$.
Note, however, that this length scale $k_F\xi^2$ is much longer than $\xi$, and the decay $r^{-1}$ is slow due to the divergence in the quasiparticle spectrum.

\section{Robustness}
\label{ref:robustness}
Let us discuss the robustness of the positive correlation at $\theta=\pi$ in a non-ideal situation.
We consider three 
types 
\begin{widetext}
\noindent
of imperfections: static fluctuations of the diameter $w$ and of the position of the emitting tip, $\delta\bm{r}$, as well as surface roughness that alters the direction of emitting electrons.
The shift of the emitting center can be accounted for by replacing the tunneling matrix elements
in (\ref{eqn:HT}) by 
\begin{equation}
T_{\bm{p}\bm{k}}
\ \to\ %
T_{\bm{p}\bm{k}}e^{-i(\bm{p}-\bm{k})\cdot\delta\bm{r}},
\label{eqn:ShiftedT}
\end{equation}
while the surface roughness would be represented by the deviation of the angle $\theta$ from $\pi$, i.e.\ $\delta\theta=\pi-\theta$.
Then, the positive peak at $\theta=\pi$ is given, instead of (\ref{eqn:Peak}), by (Appendix \ref{app:Nonideal})
\begin{align}
\delta\tilde{Q}_\text{peak}
\simeq{}&\frac{\pi^2}{32K_1^2(|\Delta|/E_C)}
\int dw\int d^3(\delta\bm{r})\int d(\delta\theta)\,
P(w,\delta\bm{r},\delta\theta)
\nonumber
\\
&\qquad\qquad\quad
{}\times
\e^{-k_F^2w^2(\delta\theta)^2/2}
\,\biggl|
e^{-(\delta r_\parallel)^2/\pi^2\xi^2}
\e^{w^2(\delta\theta)^2/8\pi^2\xi^2}
H_0^{(2)}\bm{(}
[ iw^2- iw^2(\delta\theta)^2/4+ i(\delta r_\parallel)^2-r/2k_F]/\pi^2\xi^2
\bm{)}
\nonumber
\\
&\hspace*{103truemm}
{}-\frac{4\Lambda/\pi}{\sqrt{ i r/k_Fw^2}}
\e^{ i r/2\pi^2k_F\xi^2}\e^{ i k_F(\delta r_\parallel)^2/r}
\biggr|^2,
\label{eqn:PositiveCorrNonideal}
\end{align}
\end{widetext}
where $P(w,\delta\bm{r},\delta\theta)$ is a probability distribution function that characterizes the fluctuations and $\delta r_\parallel=\hat{\bm{r}}_1\cdot\delta\bm{r}$ is the shift of the emitting center parallel to the directions to the detectors $\hat{\bm{r}}_1=-\hat{\bm{r}}_2$.
If the emitting center shifts toward one of the two detectors, coincident detections at the same distance do not imply that the two detected electrons originate from the same point, from a common Cooper pair, and the correlation is reduced when the shift is larger than the size of a Cooper pair.
That is why $\delta r_\parallel/\xi$ appears in the formula.

When $w\ll\xi$ (which is one of the requirements for a strong correlation), the first of the two contributions in the absolute value is dominant over the other, as mentioned below (\ref{eqn:Peak}).
Notice that the fluctuations in $w$ and $\delta\bm{r}$ appear there through the ratios $\delta w^2/\xi^2$ and $(\delta r_\parallel)^2/\xi^2$.
The positive correlation is hence affected by these fluctuations only when they become of order of $\xi$; otherwise, it is robust against them.
The surface roughness on the other hand shrinks the height of the positive peak mainly by the exponential factor $e^{-k_F^2w^2(\delta\theta)^2/2}$, when $\delta\theta\gtrsim 1/k_Fw$.
The positive correlation is therefore tolerant of the surface roughness up to an order of $1/\delta k=1/k_F\,\delta\theta\simeq w$.

\section{Summary}
\label{sec:Summary}
We have fully analyzed the correlations, both in space and time, of the electrons field-emitted from a superconducting tip into vacuum.
The superconductivity of the emitter leads to positive correlation between the electrons emitted in opposite directions $\theta\sim\pi$.
They can be entangled and eventually violate Bell's inequality.
A coincidence experiment can directly capture these features.
Notice that, in contrast to transport setups, one need not argue how to separate the entangled pair.\cite{note:Separate}
This is a preferable feature, as it makes nonlocality tests easier and shows that superconducting nanotips are good sources of entanglement.
Furthermore, these electron pairs are available in vacuum.

The Andreev emission plays a crucial role.
Our analytical formulas show that positive correlations and nonlocality are controlled by the parameters $w^2/\xi^2$, $r/k_F\xi^2$, and $\Delta/E_C$.
The conditions $w^2/\xi^2\lesssim1$ and $r/k_F\xi^2\lesssim1$ ensure that the paired electrons originate from a common Cooper pair, and a larger $\Delta/E_C$ makes the contribution of the Andreev emission more dominant, resulting in enhancement of the correlations.
Additional requirements, implicitly suggested and/or implied by our analysis, would help to enhance these effects and make these correlations more manifest and easily observable.
For instance, since the Andreev emission is the only process that has a nonzero emission spectrum in the ``gap,'' a better correlation would be extracted by selecting energies close to the Fermi level.

Even if entanglement were not observed, the detection of the positive correlation would still be a very challenging task.
This peculiar correlation is a direct manifestation of the singlet spin state of a Cooper pair and would be a nice probe of the symmetry of the electron pairs in a superconductor: if applied to a triplet superconductor, for instance, one should find a negative correlation.
The present setup, field emission into vacuum, would also be useful for detecting the anisotropy of unconventional superconductors.
Extension of the present analysis to more general superconductors is an interesting future subject.

\acknowledgments
We thank B. Cho, P. Facchi, R. Fazio, S. Kawabata, H. Nakazato, I. Ohba, C. Oshima, S. Pascazio, F. Taddei, and S. Tasaki for discussions.
This work is supported by the bilateral Italian-Japanese Projects II04C1AF4E on ``Quantum Information, Computation and Communication'' of the Italian Ministry of Education, University and Research, by the Joint Italian-Japanese Laboratory on ``Quantum Information and Computation'' of the Italian Ministry for Foreign Affairs, and by a Special Coordination Fund for Promoting Science and Technology and the Grant-in-Aid for Young Scientists (B) (No.\ 21740294) both from the Ministry of Education, Culture, Sports, Science and Technology, Japan.

\appendix
\section{Green's Function up to the Second Order}
\label{app:G}
Let us carry out the inverse Laplace transformation (\ref{eqn:Gdef}) to obtain $\mathcal{G}_{\bm{p}\bm{p}'}(t)$ up to second order in $\lambda$.

First, we need to invert the ``matrix'' in (\ref{eqn:Ginv}), which is formally done as
\begin{equation}
\hat{\mathcal{G}}(s)
=[\hat{\mathcal{D}}(s)
+\lambda^2\hat{\mathcal{K}}(s)]^{-1}
=\hat{\mathcal{D}}^{-1}(s)[1+\lambda^2\hat{\mathcal{K}}(s)\hat{\mathcal{D}}^{-1}(s)]^{-1},
\label{eqn:Gformal}
\end{equation}
where $\hat{\mathcal{D}}_{\bm{p}\bm{p}'}(s)=(s+ i\mathcal{E}_p)\delta^3(\bm{p}-\bm{p}')$.
Notice here that it is possible to show that
\begin{equation}
\Det[1+\lambda^2\hat{\mathcal{K}}(s)\hat{\mathcal{D}}^{-1}(s)]\neq0
\end{equation}
for a sufficiently small $\lambda$.\cite{ref:LateralEffects}
Indeed, the determinant is evaluated as
\begin{align}
&\Det[1+\lambda^2\hat{\mathcal{K}}(s)\hat{\mathcal{D}}^{-1}(s)]
=\e^{\Tr\log[1+\lambda^2\hat{\mathcal{K}}(s)\hat{\mathcal{D}}^{-1}(s)]},
\intertext{which reads, in the weak-coupling regime,}
&\ %
=1+\lambda^2\Tr[\hat{\mathcal{K}}(s)\hat{\mathcal{D}}^{-1}(s)]+O(\lambda^4)
\nonumber
\displaybreak[0]
\\
&\ %
=1+\lambda^2\int\d^3\bm{p}\tr\!\left(
\hat{\mathcal{K}}_{\bm{p}\bm{p}}(s)\frac{1}{s+ i\mathcal{E}_p}
\right)+O(\lambda^4)
\nonumber
\displaybreak[0]\\
&\ %
=1+\lambda^2\int\d^3\bm{p}\int\d^3\bm{k}
\,\biggl(
\frac{|T_{\bm{p}\bm{k}}u_k|^2}{(s+ i\varepsilon_p)(s+ i\omega_k)}
\nonumber
\\
&\ \qquad\qquad\qquad\qquad\qquad\quad
{}+\frac{|T_{\bm{p}\bm{k}}u_k|^2}{(s- i\varepsilon_p)(s- i\omega_k)}
\biggr)
\nonumber
\displaybreak[0]
\\
&\ \quad
{}+\lambda^2\int\d^3\bm{p}\int\d^3\bm{k}
\,\biggl(
\frac{|T_{\bm{p}\bm{k}}v_k|^2}{(s+ i\varepsilon_p)(s- i\omega_k)}
\nonumber
\\
&\quad\qquad\qquad\qquad\qquad\quad
{}+\frac{|T_{\bm{p}\bm{k}}v_k|^2}{(s- i\varepsilon_p)(s+ i\omega_k)}
\biggr)+O(\lambda^4).
\label{eqn:DetG}
\end{align}
For a regular tunneling matrix $T_{\bm{p}\bm{k}}$ [i.e., if the spectral functions  $J_u(\varepsilon,\omega)=\int\d^3\bm{p}\int\d^3\bm{k}\,|T_{\bm{p}\bm{k}}u_k|^2\delta(\varepsilon_p-\varepsilon)\delta(\omega_k-\omega)$ and $J_v(\varepsilon,\omega)$ have good spectral properties: free from a nonlocal spectrum like $\delta(\varepsilon-\omega)$ and vanishing for $\varepsilon,\omega\to-\mu,\infty$], the integral in (\ref{eqn:DetG}) is bounded for any $s$ (even on the imaginary axis of the complex $s$ plane), and the determinant (\ref{eqn:DetG}) can always be made non-vanishing by choosing a sufficiently small $\lambda$.
It is therefore possible to expand the second factor in the right-hand side of (\ref{eqn:Gformal}) in a power series in $\lambda^2$,
\begin{equation}
\hat{\mathcal{G}}(s)
=\hat{\mathcal{D}}^{-1}(s)[1-\lambda^2\hat{\mathcal{K}}(s)\hat{\mathcal{D}}^{-1}(s)+O(\lambda^4)],
\end{equation}
i.e.,
\begin{align}
\hat{\mathcal{G}}_{\bm{p}\bm{p}'}(s)
={}&\frac{1}{s+ i\mathcal{E}_p}\delta^3(\bm{p}-\bm{p}')
\nonumber
\displaybreak[0]
\\
&{}-\lambda^2\frac{1}{s+ i\mathcal{E}_p}
\hat{\mathcal{K}}_{\bm{p}\bm{p}'}(s)
\frac{1}{s+ i\mathcal{E}_{p'}}
+O(\lambda^4).
\end{align}
Its inverse Laplace transformation yields $\mathcal{G}_{\bm{p}\bm{p}'}(t)$ up to second order in $\lambda$,
\begin{align}
&\mathcal{G}_{\bm{p}\bm{p}'}(t)
=\e^{- i\mathcal{E}_pt}\delta^3(\bm{p}-\bm{p}')
\nonumber
\\
&\qquad
{}-\lambda^2\int_{C_B}\frac{\d s}{2\pi i }
\begin{pmatrix}
\medskip
\frac{\hat{K}_{\bm{p}\bm{p}'}^{11}(s)}{(s+ i\varepsilon_p)(s+ i\varepsilon_{p'})}
&\frac{\hat{K}_{\bm{p}\bm{p}'}^{12}(s)}{(s+ i\varepsilon_p)(s- i\varepsilon_{p'})}
\\
\frac{\hat{K}_{\bm{p}\bm{p}'}^{21}(s)}{(s- i\varepsilon_p)(s+ i\varepsilon_{p'})}
&\frac{\hat{K}_{\bm{p}\bm{p}'}^{22}(s)}{(s- i\varepsilon_p)(s- i\varepsilon_{p'})}
\end{pmatrix}
\e^{st}
\nonumber
\\
&\hspace*{70mm}
{}+O(\lambda^4).
\end{align}

Let us take the stationary limit $t\to\infty$ in the interaction picture $\bar{\mathcal{G}}_{\bm{p}\bm{p}'}(t)
=\mathcal{G}_{\bm{p}\bm{p}'}(t)\e^{ i\mathcal{E}_{p'}t_0}$.
It proceeds, for instance, as follows:
by noting
\begin{align}
&\frac{1}{(s+ i\varepsilon_p)(s- i\varepsilon_{p'})}
\nonumber
\\
&\qquad
=\frac{1}{ i(\varepsilon_p+\varepsilon_{p'})\pm0^+}
\left(
\frac{1}{s- i\varepsilon_{p'}}
-\frac{1}{s+ i\varepsilon_p}
\right),
\end{align}
one gets, for $t,t_0\to\infty$ keeping $t-t_0$ finite,
\begin{align}
&\int_{C_B}\frac{\d s}{2\pi i }
\frac{\hat{K}_{\bm{p}\bm{p}'}^{12}(s)}{(s+ i\varepsilon_p)(s- i\varepsilon_{p'})}
\e^{st}\e^{- i\varepsilon_{p'}t_0}
\nonumber
\displaybreak[0]
\\
&\ \ %
=\int_{C_B}\frac{\d s}{2\pi i }
\hat{K}_{\bm{p}\bm{p}'}^{12}(s)\,\biggl(
\frac{1}{ i(\varepsilon_p+\varepsilon_{p'})+0^+}
\frac{\e^{(s- i\varepsilon_{p'})t}}{s- i\varepsilon_{p'}}
\nonumber
\displaybreak[0]
\\
&\quad\quad\qquad\qquad\quad\ %
{}-\frac{\e^{- i(\varepsilon_p+\varepsilon_{p'})t}}{ i(\varepsilon_p+\varepsilon_{p'})+0^+}
\frac{\e^{(s+ i\varepsilon_p)t}}{s+ i\varepsilon_p}
\biggr)\,
\e^{ i\varepsilon_{p'}(t-t_0)}
\nonumber
\displaybreak[0]
\\
&\ \ %
\to\frac{\hat{K}_{\bm{p}\bm{p}'}^{12}( i\varepsilon_{p'}+0^+)}{ i(\varepsilon_p+\varepsilon_{p'})+0^+}\e^{ i\varepsilon_{p'}(t-t_0)}.
\end{align}
Recall the formula
\begin{equation}
\lim_{t\to\infty}\frac{\e^{- i xt}}{x\pm i0^+}
=\begin{cases}
\medskip
\displaystyle
-2\pi i\,\delta(x),\\
\displaystyle
0.
\end{cases}
\label{eqn:ExpScatPole}
\end{equation}
Similar treatments are applied to the other components, and one ends up with (\ref{eqn:Gasymp}).

\section{Far-Field Spectra}
\label{app:AsympWave}
Let us sketch the derivations of the energy spectra in the far-field limit.
Let us first look at the asymptotic behavior of the wave functions in (\ref{eqn:WaveFunction}).\cite{ref:LateralEffects}
At far places from the emitter, $k_Fr\gg1$, only momenta oriented along $\pm\hat{\bm{r}}_i$ ($i=1,2$) contribute to the propagation of the emitted electrons.
The saddle-point approximation (method of steepest descent) for the integrations over the orientations of the momenta yields
\begin{align}
&\int\frac{\d^3\bm{p}}{\sqrt{(2\pi)^3}\,i}
\frac{T_{\bm{p}\bm{k}}}{\varepsilon_p\mp\omega_k-i0^+}
\e^{i\bm{p}\cdot\bm{r}}
\nonumber
\displaybreak[0]
\\
&\quad
\simeq
-\frac{1}{\sqrt{2\pi}\,r}
\int_0^\infty\d p\,p\,\biggl(
\frac{T_{(p\hat{\bm{r}})\bm{k}}}{\varepsilon_p\mp\omega_k-i0^+}\e^{i pr}
\nonumber
\\
&\quad\qquad\qquad\qquad\qquad\qquad
-\frac{T_{(-p\hat{\bm{r}})\bm{k}}}{\varepsilon_p\mp\omega_k-i0^+}\e^{-i pr}
\biggr)
\nonumber
\displaybreak[0]
\\
&\quad
=-\frac{1}{\sqrt{2\pi}\,r}
\int_{-\infty}^\infty\d p\,p
\frac{T_{(p\hat{\bm{r}})\bm{k}}}{\varepsilon_p\mp\omega_k-i0^+}\e^{i pr}\nonumber
\displaybreak[0]
\\
&\quad
\simeq m\sqrt{2\pi}\,\theta(\mu\pm\omega_k)
T_{(p_\pm\hat{\bm{r}})\bm{k}}
\frac{\e^{ ip_\pm r}}{ i r},
\label{eqn:AsympWave}
\end{align}
where $p_\pm$ are defined in (\ref{eqn:Ppm}) and formula (\ref{eqn:ExpScatPole}) is used.
One therefore gets (\ref{eqn:AsympWaveFunction}).
By plugging these expressions together with the tunneling matrix (\ref{eqn:EmissionMatrix}) and (\ref{eqn:Setup}) into the correlation functions (\ref{eqn:Gamma}) and (\ref{eqn:ChiTh}), and by performing the Fourier transformation with respect to time $\tau=t_1-t_2$, the energy spectra of the correlation functions in the NESS and in the far-field limit, (\ref{eqn:GammaE}) and (\ref{eqn:ChiThE}) with (\ref{eqn:Z}) and (\ref{eqn:Zth}), are obtained.

Similar treatment to (\ref{eqn:AsympWave}) applies to (\ref{eqn:Andreev}), yielding (for $r_1=r_2=r$)
\begin{widetext}
\begin{align}
&\chi_0(\bm{r}_1,t_1;\bm{r}_2,t_2)
\nonumber\displaybreak[0]\\
&\quad
\simeq-\frac{\lambda^2}{2\pi r^2}
\int\d^3\bm{k}\,u_kv_k
\int_{-\infty}^\infty\d p_1\,p_1
\int_{-\infty}^\infty\d p_2\,p_2
\frac{T_{(p_1\hat{\bm{r}}_1)\bm{k}}T_{(p_2\hat{\bm{r}}_2)(-\bm{k})}
}{\varepsilon_{p_1}+\varepsilon_{p_2}-i0^+}
\left(
\frac{1}{\varepsilon_{p_1}+\omega_k-i0^+}
+\frac{1}{\varepsilon_{p_2}+\omega_k-i0^+}
\right)
\nonumber\displaybreak[0]\\
&\hspace*{130mm}
{}\times
\e^{ i(p_1+p_2)r}
\e^{- i\varepsilon_{p_1}(t_1-t_2)}
\nonumber\displaybreak[0]\\
&\quad\quad
{}+\frac{\lambda^2}{2\pi r^2}
\int\d^3\bm{k}\,u_kv_k
\int_{-\infty}^\infty\d p_1\,p_1
\int_{-\infty}^\infty\d p_2\,p_2
\frac{T_{(p_1\hat{\bm{r}}_1)\bm{k}}T_{(p_2\hat{\bm{r}}_2)(-\bm{k})}\e^{ i(p_1+p_2)r}}{(\varepsilon_{p_1}-\omega_k-i0^+)(\varepsilon_{p_2}+\omega_k-i0^+)}
(\e^{- i\varepsilon_{p_1}(t_1-t_2)}-\e^{- i\omega_k(t_1-t_2)}).
\end{align}
In the second term, the poles of the denominators (on-shell contributions) are significant for large $k_Fr$ [see formula (\ref{eqn:ExpScatPole})] but such contributions are suppressed by the last factor.
The first term is therefore responsible for the far fields of the emitted couple.
The pole of its first denominator (which describes the propagation of the emitted pair in the vacuum) yields
\begin{equation}
\chi_0(\bm{r}_1,t_1;\bm{r}_2,t_2)
\simeq\frac{\lambda^2m}{ i r^2}
\int\d^3\bm{k}\,u_kv_k
\int_{-\infty}^\infty\d p\,p
\frac{\theta(\mu-\varepsilon_p)}{\varepsilon_p+\omega_k-i0^+}
T_{(p\hat{\bm{r}}_1)\bm{k}}T_{(\bar{p}\hat{\bm{r}}_2)(-\bm{k})}
\e^{ i(p+\bar{p})r}\e^{- i\varepsilon_p(t_1-t_2)}
+(1\leftrightarrow2),
\end{equation}
where $\bar{p}=p(-\varepsilon_p)$ with $p(E)$ defined in (\ref{eqn:Ppm}).
Its Fourier transformation with respect to $\tau=t_1-t_2$ gives the spectral representation
\begin{align}
\mathcal{X}_0(\bm{r}_1,\bm{r}_2;E)
={}&{-\frac{2\pi\lambda^2m^2}{ i r^2}}
\theta(\mu-|E|)
\int\d^3\bm{k}\,u_kv_k
\frac{2\omega_k}{E^2-\omega_k^2+i0^+}
T_{[p(-E)\hat{\bm{r}}_1]\bm{k}}T_{[p(E)\hat{\bm{r}}_2](-\bm{k})}
\e^{ i[p(E)+p(-E)]r}
\nonumber
\displaybreak[0]
\\
&{}+\frac{2\pi\lambda^2m^2}{ i r^2}
\theta(\mu-|E|)
\int\d^3\bm{k}\,u_kv_k
\frac{1}{E-\omega_k+i0^+}
T_{[-p(-E)\hat{\bm{r}}_1]\bm{k}}T_{[p(E)\hat{\bm{r}}_2](-\bm{k})}
\e^{ i[p(E)-p(-E)]r}
\nonumber
\displaybreak[0]
\\
&{}-\frac{2\pi\lambda^2m^2}{ i r^2}
\theta(\mu-|E|)
\int\d^3\bm{k}\,u_kv_k
\frac{1}{E+\omega_k-i0^+}
T_{[p(-E)\hat{\bm{r}}_1](-\bm{k})}T_{[-p(E)\hat{\bm{r}}_2]\bm{k}}
\e^{- i[p(E)-p(-E)]r}.
\intertext{Only the first term survives for large $k_Fr$, due to formula (\ref{eqn:ExpScatPole}),}
\simeq{}&{-\frac{2\pi m^2\lambda^2}{ i r^2}}
\theta(\mu-|E|)
\int\d^3\bm{k}\,u_kv_k
\frac{2\omega_k}{E^2-\omega_k^2+i0^+}
T_{[p(-E)\hat{\bm{r}}_1]\bm{k}}T_{[p(E)\hat{\bm{r}}_2](-\bm{k})}
\e^{ i[p(E)+p(-E)]r}.
\end{align}
\end{widetext}
For the tunneling matrix (\ref{eqn:EmissionMatrix}) with the spherical setup (\ref{eqn:Setup}), this yields (\ref{eqn:Chi0E}) with (\ref{eqn:Z0}), by noting the formulas for the error functions, (\ref{eqn:erfccut}) and (\ref{eqn:erfi}).

\section{Andreev Emission}
\label{app:Andreev}
We briefly recapitulate the field-theoretical description of the Andreev emission process via a wave operator.
The 
wave 
operator $\mathcal{W}$ is the operator that describes the scattering of an initial eigenstate of $H_0$ into a scattering state 
by a scattering Hamiltonian $H$.
It is formally defined by
\begin{equation}
\mathcal{W}=\lim_{t\to\infty}\e^{- iHt}\e^{ iH_0t},\quad
H=H_0+V.
\end{equation}
Let $\ket{E}$ be the eigenstate 
of $H_0$ belonging to its eigenvalue $E$ and suppose that the Hamiltonian $H$ does not have 
any 
bound state.
Then, the wave operator $\mathcal{W}$ is given by
\begin{equation}
\mathcal{W}\ket{E}
=\left(
1+\frac{1}{E-H+i0^+}V
\right)\ket{E}.
\label{eqn:LippmannSchwinger}
\end{equation}
Assume further that $H_0$ admits only one discrete eigenvalue $E_0$ besides a continuous spectrum $E$ (as is often the case in field theory; the vacuum state), and
\begin{equation}
\bra{E_0}V\ket{E_0}=0.
\end{equation}
Equation (\ref{eqn:LippmannSchwinger}) is valid also for such a discrete spectrum of $H_0$, and the matrix element for the scattering of $\ket{E_0}$ into $\ket{E}$ is described by
\begin{widetext}
\begin{equation}
\bra{E}\mathcal{W}\ket{E_0}
=\frac{1}{E_0-E+i0^+}\,\biggl(
\bra{E}V\ket{E_0}
+\bra{E}V\frac{1}{E_0-H_0+i0^+}V\ket{E_0}
+\bra{E}V\frac{1}{E_0-H+i0^+}V\frac{1}{E_0-H_0+i0^+}V\ket{E_0}
\biggr).
\label{eqn:Wdisc2cont}
\end{equation}
\end{widetext}

Let us apply this formalism to the present case, i.e.\ to the emission of electrons from a superconductor.
The Hamiltonian (\ref{eqn:HamiltonianChemical}) meets the above requirements.
In particular, the only discrete state is the ground state, vacuum outside and the BCS state inside the emitter,
\begin{equation}
\ket{0}=\ket{0}_V\otimes\ket{\text{BCS}}_S,
\end{equation}
with $E_0=0$.
We are interested in the emission of an electron pair from this ground state, in the absence of the quasiparticle excitations in the emitter.
Hence, the relevant matrix element is
\begin{equation}
\bra{\bm{p}_2s_2;\bm{p}_1s_1}\mathcal{W}\ket{0},
\quad
\ket{\bm{p}_2s_2;\bm{p}_1s_1}
=c_{\bm{p}_2s_2}^\dag c_{\bm{p}_1s_1}^\dag\ket{0},
\end{equation}
where, in the final state, two electrons are found outside while the emitter remains in the BCS state.
The lowest nontrivial contribution to this matrix element appears at the second order in $H_T$ given in (\ref{eqn:HT}) and reads \cite{ref:HekkingNazarov,
ref:Loss,
ref:Prada}
\begin{widetext}
\begin{align}
\bra{\bm{p}_2s_2;\bm{p}_1s_1}\mathcal{W}\ket{0}
&=\lambda^2
\frac{1}{\varepsilon_{p_1}+\varepsilon_{p_2}-i0^+}
\bra{\bm{p}_2s_2;\bm{p}_1s_1}
H_T\frac{1}{\mathcal{H}_0-i0^+}H_T
\ket{0}
+O(\lambda^4)
\nonumber
\displaybreak[0]
\\
&=\lambda^2(\delta_{s_1{\uparrow}}\delta_{s_2{\downarrow}}
-\delta_{s_1{\downarrow}}\delta_{s_2{\uparrow}})
\int\d^3\bm{k}\,u_kv_k
\frac{T_{\bm{p}_1\bm{k}}T_{\bm{p}_2(-\bm{k})}}{\varepsilon_{p_1}+\varepsilon_{p_2}-i0^+}\left(
\frac{1}{\varepsilon_{p_1}+\omega_k-i0^+}
+\frac{1}{\varepsilon_{p_2}+\omega_k-i0^+}
\right)+O(\lambda^4).
\label{eqn:AndreevAmp}
\end{align}
\end{widetext}
Clearly, the emitted pair of electrons is in the singlet state, and the amplitude is symmetric under exchange between $p_1$ and $p_2$.
In principle, one can compute any higher-order processes according to (\ref{eqn:LippmannSchwinger}) and (\ref{eqn:Wdisc2cont}).
It is interesting to note that virtual processes (propagators between $H_T$'s) are involved in the Andreev process.

\section{Integral Formulas}
\label{app:IntForm}
It is useful to quote some integral formulas used in the derivation of some analytical results. Some of the equations that follow require extensions of known formulas.

\subsection{An Integral with the Fermi Distribution Function}
An integral with the Fermi distribution function:\cite{ref:IntegralTable}
\begin{equation}
\int_{-\infty}^\infty\d x\,\frac{\e^{\alpha x}}{\e^{\beta x}+1}
=\frac{\pi/\beta}{\sin(\pi\alpha/\beta)}
\quad(\Re\beta>\Re\alpha>0).
\label{eqn:IntForm}
\end{equation}

\subsection{Bessel Functions}
The density of state of the superconductor, which diverges at the gap with $1/\sqrt{E^2-|\Delta|^2}$, yields Bessel functions,
\begin{subequations}
\begin{multline}
J_\nu(z)
=\frac{2^{1-\nu}z^\nu}{\sqrt{\pi}\,\Gamma(\nu+1/2)}
\int_0^1\d t\,(1-t^2)^{\nu-1/2}\cos zt
\\
(\Re\nu>-1/2),
\end{multline}
\begin{multline}
Y_\nu(x)
=-\frac{2^{\nu+1}x^{-\nu}}{\sqrt{\pi}\,\Gamma(1/2-\nu)}
\int_1^\infty\d t\,(t^2-1)^{-\nu-1/2}\cos xt
\\
(x>0,\ |\!\Re\nu|<1/2),
\label{eqn:BesselY}
\end{multline}
\begin{multline}
I_\nu(z)
=\frac{2^{1-\nu}z^\nu}{\sqrt{\pi}\,\Gamma(\nu+1/2)}
\int_0^1\d t\,(1-t^2)^{\nu-1/2}\cosh zt
\\
(\Re\nu>-1/2),
\end{multline}
\begin{multline}
K_\nu(z)
=\frac{\sqrt{\pi}\,z^\nu}{2^\nu\Gamma(\nu+1/2)}
\int_1^\infty\d t\,(t^2-1)^{\nu-1/2}\e^{-zt}
\\
(\Re z>0,\ \Re\nu>-1/2).
\label{eqn:BesselK}
\end{multline}
\end{subequations}
In particular, the following formulas are used in the text:
(i) for a real number $x$,
\begin{equation}
\int_{0+i0^+}^{\infty+i0^+}\d t\,
\frac{\cos xt}{\sqrt{t^2-1}}
=\frac{\pi}{2i}H_0^{(2)}(|x|),
\end{equation}
where
\begin{equation}
\begin{cases}
H_\nu^{(1)}(z)=J_\nu(z)+ i Y_\nu(z),\\
H_\nu^{(2)}(z)=J_\nu(z)- i Y_\nu(z)
\end{cases}
\end{equation}
are Hankel functions;
(ii) for $\Re z>0$,
\begin{align}
&\int_{0+i0^+}^{\infty+i0^+}\d t\,
\frac{\e^{-zt^2}}{\sqrt{t^2-1}}
\nonumber
\\
&\quad
=-\frac{i}{2}\int_0^1\d u\,\frac{\e^{-zu}}{\sqrt{u(1-u)}}
+\frac{1}{2}\int_1^\infty\d u\,\frac{\e^{-zu}}{\sqrt{u(u-1)}}
\nonumber
\displaybreak[0]
\\
&\quad
=-\frac{i}{2}\e^{-z/2}
\int_{-1}^1\d t\,\frac{\e^{-zt/2}}{\sqrt{1-t^2}}
+\frac{1}{2}\e^{-z/2}
\int_1^\infty\d t\,\frac{\e^{-zt/2}}{\sqrt{t^2-1}}
\nonumber
\displaybreak[0]
\\
&\quad
=\frac{1}{2}\e^{-z/2}\Bigl(
K_0(z/2)-\pi iI_0(z/2)
\Bigr)
\nonumber
\displaybreak[0]
\\
&\quad
=\frac{\pi}{4i }\e^{-z/2}H_0^{(2)}( i z/2)
\qquad(\Re z>0),
\end{align}
by noting the relations
\begin{equation}
I_\nu(z)=\e^{-\nu\pi i/2}J_\nu( i z),\ \ %
K_\nu(z)=\frac{\pi}{2}\e^{(\nu+1)\pi i/2}H_\nu^{(1)}( i z).
\end{equation}

The asymptotic behavior of some of the Bessel functions is useful for the discussion in the text:
\begin{align}
H_\nu^{(1)}(x)
&=[H_\nu^{(2)}(x)]^*
\sim\sqrt{\frac{2}{\pi x}}
\e^{ i[x-(2\nu+1)\pi/4]}
\quad(x\to\infty),
\label{eqn:BesselHRealAsymp}
\displaybreak[0]\\
H_\nu^{(1)}( i z)
&\sim\sqrt{\frac{2}{\pi i z}}
\e^{-z-(2\nu+1)\pi i/4}
\quad(\Re z>0,\ |z|\to\infty),
\displaybreak[0]\\
H_\nu^{(2)}( i z)
&\sim
\sqrt{\frac{2}{\pi i z}}
\,\Bigl(
\e^{z+(2\nu+1)\pi i/4}
\nonumber\\[-2mm]
&\qquad\qquad\ \ %
{}-\theta(\Im z)(1+\e^{2\nu\pi i})\e^{-z-(2\nu+1)\pi i/4}
\Bigr)
\nonumber\\
&\qquad\qquad\qquad\qquad\qquad
(\Re z>0,\ \Im z\to\pm\infty),
\label{eqn:BesselHCompAsymp}
\displaybreak[0]\\
K_\nu(z)
&\sim
\begin{cases}
\medskip
\displaystyle
\sqrt{\frac{\pi}{2z}}\e^{-z}
&(\Re z>0,\ |z|\to\infty),\\
\displaystyle
\frac{1}{2}\Gamma(\nu)\left(\frac{x}{2}\right)^{-\nu}
&(\nu>0,\ x\to0^+).
\end{cases}
\label{eqn:BesselKAsymp}
\end{align}
Note also the fundamental relation
\begin{equation}
x^\nu\frac{\d}{\d x}x^{-\nu}K_\nu(x)
=-K_{\nu+1}(x).
\label{eqn:LadderBesselK}
\end{equation}

\subsection{Error Functions}
The complementary error function $\erfc(z)$ 
is an entire function with no branch cut and endowed with a property
\begin{equation}
\erfc(-z)=2-\erfc(z).
\end{equation}
For $\Re z>0$, the following expression is available:
\begin{equation}
\erfc(z)
=\pm\frac{\e^{-z^2}}{\pi i}\int_{-\infty}^\infty\d t\,
\frac{\e^{-t^2}}{t\mp i z}
\quad(\Re z>0).
\label{eqn:erfccut}
\end{equation}
The complementary error function is related to the imaginary error function by
\begin{equation}
\erfc(z)=1+ i\erfi( i z),\quad
\erfi(-z)=-\erfi(z),
\label{eqn:erfi}
\end{equation}
which asymptotically behaves as
\begin{equation}
\erfi(z)\sim i\epsilon(\Im z)+\frac{\e^{z^2}}{\sqrt{\pi}\,z}\left[
1+O\!\left(\frac{1}{z^2}\right)
\right]\quad(|z|\to\infty)
\end{equation}
with the convention $\epsilon(0)=0$ for the sign function.

\section{Correlation Functions with a Shifted Emitting Center}
\label{app:Nonideal}
If the emitting center is displaced in space by a vector $\delta\bm{r}$, the transmission matrix $T_{\bm{p}\bm{k}}$ acquires a phase like (\ref{eqn:ShiftedT}).
In the far-field regime $k_Fr\gg1$ with a small displacement $\delta r\ll r$, this just induces simple phases in the spectra $\Gamma$, $\mathcal{X}_\text{th}$, and $\mathcal{X}_0$ in (\ref{eqn:Spectra}) of the two-point correlation functions $\gamma$, $\chi_\text{th}$, and $\chi_0$:
\begin{widetext}
\begin{subequations}
\begin{align}
\gamma(\bm{r}_1,t_1;\bm{r}_2,t_2)
\ \to\ \tilde{\gamma}(\bm{r}_1,t_1;\bm{r}_2,t_2)
&=\int_{-\infty}^\infty\frac{\d E}{2\pi}\,
\e^{ ip(E)(\hat{\bm{r}}_1-\hat{\bm{r}}_2)\cdot\bm\delta\bm{r}}\Gamma(\bm{r}_1,\bm{r}_2;E)\e^{ i E(t_1-t_2)},
\displaybreak[0]\\
\chi_\text{th}(\bm{r}_1,t_1;\bm{r}_2,t_2)
\ \to\ \tilde{\chi}_\text{th}(\bm{r}_1,t_1;\bm{r}_2,t_2)
&=\int_{-\infty}^\infty\frac{\d E}{2\pi}\,
\e^{- i[p(-E)\hat{\bm{r}}_1+p(E)\hat{\bm{r}}_2]\cdot\bm\delta\bm{r}}\mathcal{X}_\text{th}(\bm{r}_1,\bm{r}_2;E)\e^{ i E(t_1-t_2)},
\end{align}
\end{subequations}
and a similar modification for $\chi_0$, where $p(E)$ is defined in (\ref{eqn:Ppm}).

The same treatment as the one for $\gamma$ yields
\begin{equation}
\tilde{\gamma}(\bm{r}_1,t_1;\bm{r}_2,t_2)
=\e^{ i k_F(\hat{\bm{r}}_1-\hat{\bm{r}}_2)\cdot\bm\delta\bm{r}} 
\gamma(\bm{r}_1,t_1;\bm{r}_2,t_2),
\label{eqn:GammaNonideal}
\end{equation}
while $\tilde{\chi}_\text{th/0}$ is estimated to be $\chi_\text{th/0}$ with substitutions $r\to r-(\hat{\bm{r}}_1+\hat{\bm{r}}_2)\cdot\delta\bm{r}/2$ and $t_1-t_2\to t_1-t_2+k_F(\hat{\bm{r}}_1-\hat{\bm{r}}_2)\cdot\delta\bm{r}/2\mu$.
For the coincident detections $t_1-t_2=0$ with $k_F(\delta r)^2/r\ll1$, the integrals involved in $\tilde{\chi}_\text{th/0}$ are estimated as [defining $a=k_F(\hat{\bm{r}}_1-\hat{\bm{r}}_2)\cdot\delta\bm{r}/2\mu$]
\begin{align}
\int_{-\infty}^\infty\d E\,\frac{\Delta}{\sqrt{E^2-|\Delta|^2}}\e^{- i E^2/2\kappa_\pm}\e^{ i Ea}
&
\simeq\int_{-\infty}^\infty\d E\,\frac{\Delta}{\sqrt{E^2-|\Delta|^2}}
\left(
1-\frac{1}{2}E^2a^2+\cdots
\right)
\e^{- i E^2/2\kappa_\pm}
\nonumber\\
&
\simeq\int_{-\infty}^\infty\d E\,\frac{\Delta}{\sqrt{E^2-|\Delta|^2}}
\e^{- i E^2/2\kappa_\pm}\e^{-E^2a^2/2}
\nonumber\\
&
=\frac{\pi}{4i }
\Delta
\e^{- i|\Delta|^2/4\kappa_\pm}
\e^{-|\Delta|^2a^2/4}
H_0^{(2)}( i|\Delta|^2/4\kappa_\pm+|\Delta|^2a^2/4),
\end{align}
and one ends up with 
\begin{align}
\tilde{\chi}_0(\bm{r}_1,t;\bm{r}_2,t)
\simeq\frac{\pi A\Delta\e^{2i k_Fr}}{4i \sin(\theta/2)}
\,\Biggl(
\e^{-4k_F^2w^2\sin^2[(\pi-\theta)/4]}
&\e^{- i r/2\pi^2k_F\xi^2}
\e^{-(w^2/\pi^2\xi^2)[\nu-\sin(\theta/2)]}
\e^{-[(\hat{\bm{r}}_1-\hat{\bm{r}}_2)\cdot\delta\bm{r}]^2/4\pi^2\xi^2}
\nonumber\\[-2truemm]
&{}\times
H_0^{(2)}\bm{(}
\{ i\nu w^2+ i[(\hat{\bm{r}}_1-\hat{\bm{r}}_2)\cdot\delta\bm{r}]^2/4-r/2k_F\}/\pi^2\xi^2
\bm{)}
\nonumber\\
&\qquad\qquad
{}-\frac{4\Lambda/\pi}{\sqrt{ i r/k_Fw^2}}
\e^{-k_F^2w^2\cos^2(\theta/2)}
\e^{ i k_F[(\hat{\bm{r}}_1-\hat{\bm{r}}_2)\cdot\delta\bm{r}]^2/4r}
\Biggr),
\end{align}
instead of (\ref{eqn:Chi0Coinc}).
Its ratio to (\ref{eqn:GammaNonideal}) at $\theta=\pi$ gives the integrand of (\ref{eqn:PositiveCorrNonideal}).
\end{widetext}


\begin{thebibliography}{10}
\bibitem{note:Separate}
For instance, a couple of entangled electrons from a superconductor may enter the same lead, which makes it impossible to use each electron separately.
To solve this problem, various arrangements are proposed.\cite{ref:Loss,ref:Prada}
\end{thebibliography}
\end{document}